\documentclass[pre,floatfix,twocolumn,showpacs,superscriptaddress]{revtex4-2}

\usepackage{graphicx}
\usepackage{bm}

\newcommand{\ve}[1][K]{\mathbf{#1}}

\newcommand{\mx}[1]{\textcolor{black}{#1}}

\usepackage{amsmath}
\usepackage{amssymb}
\usepackage{latexsym}
\usepackage{multirow}
\usepackage{xcolor}
\usepackage{hyperref}

\begin{document}

\title{ Imperfect Narrow Escape problem}

\author{T. Gu\'erin}
\affiliation{Laboratoire Ondes et Mati\`ere d'Aquitaine, CNRS, UMR 5798, Universit\'e de Bordeaux, F-33400 Talence, France}
\author{M. Dolgushev}
\author{O. B\'enichou}
\affiliation{Sorbonne Universit\'{e}, CNRS, Laboratoire de Physique Th\'{e}orique de la Mati\`{e}re Condens\'{e}e, LPTMC, F-75005 Paris, France}
\author{R. Voituriez}
\affiliation{Sorbonne Universit\'{e}, CNRS, Laboratoire de Physique Th\'{e}orique de la Mati\`{e}re Condens\'{e}e, LPTMC, F-75005 Paris, France}
\affiliation{Sorbonne Universit\'{e}, CNRS, Laboratoire Jean Perrin, LJP, F-75005 Paris, France}

\bibliographystyle{apsrev}

\date{\today}

\begin{abstract} 
We consider the kinetics of the imperfect narrow escape problem, i.e. the time it takes for a particle diffusing in a confined medium of generic shape to reach and to be adsorbed by a small,  imperfectly reactive patch embedded in the boundary of the domain, in  two or three dimensions. Imperfect reactivity is modeled by an intrinsic surface reactivity $\kappa$ of the patch, giving rise to Robin boundary conditions. We present a formalism to calculate the exact asymptotics of the mean reaction time in the limit of large volume of the confining domain. We obtain exact explicit results in the two limits of large and small reactivities of the reactive patch, and a semi-analytical expression in the general case. Our approach reveals an anomalous scaling of  the mean reaction time as the inverse square root of the reactivity in the  large reactivity limit, valid for an initial position  near the extremity of the reactive patch.  We  compare our exact results with those obtained within the ``constant flux approximation''; we show that this approximation turns out to give exactly the next-to-leading order term of the small reactivity limit, and provides a good approximation of the reaction time far from the reactive patch for all reactivities, but not in the vicinity of the boundary of the reactive patch due to the above mentioned anomalous scaling. These results thus provide a general framework to quantify  the mean reaction times for the imperfect narrow escape problem. 
\end{abstract}

\maketitle
  
\section{Introduction}  
  
How much time does it take for a random walker to reach a target point? 
The answer to this question has received a lot of attention in the last decade in the physics literature~\cite{Redner:2001a,Condamin2007,pal2017first,grebenkov2016universal,benichou2010optimal,vaccario2015first,metzler2014first,Schuss2007,newby2016first}. First passage problems appear in various areas of biological and soft matter physics and are in particular relevant to the problem of reaction kinetics, since two reactants have to meet before being able to react~\cite{RiceBook,Berg1985}. When the reaction is ``perfect'', i.e. when it occurs instantaneously upon  each encounter, its kinetics is controlled by the first passage statistics of one reactant molecule, seen as a random walker, to the second reactant, seen as a ``target''. 
However, many reactions do not occur at first contact between the random walker and the targets, leading to imperfect reactivity.  
Imperfect reactivity can have diverse origins at the microscopic scale, such as orientational constraints on the reactive particles~\cite{Berg1985}, the fact that the surface of the reactive particles  is not entirely covered by reactive patches (such as in the chemoreception problem \cite{berg1977physics}), the need to overcome an energetic \cite{shoup1982role} (or entropic \cite{zhou1991rate}) activation barrier before reaction, the presence of a gate that can be randomly closed or opened when the reactant meets the target \cite{reingruber2009gated,benichou2000kinetics}, etc (see Ref.~\cite{grebenkov2019imperfect} and references therein for a recent review on imperfect reactivity). 

\begin{figure}[h!]
\includegraphics[width=8cm,clip]{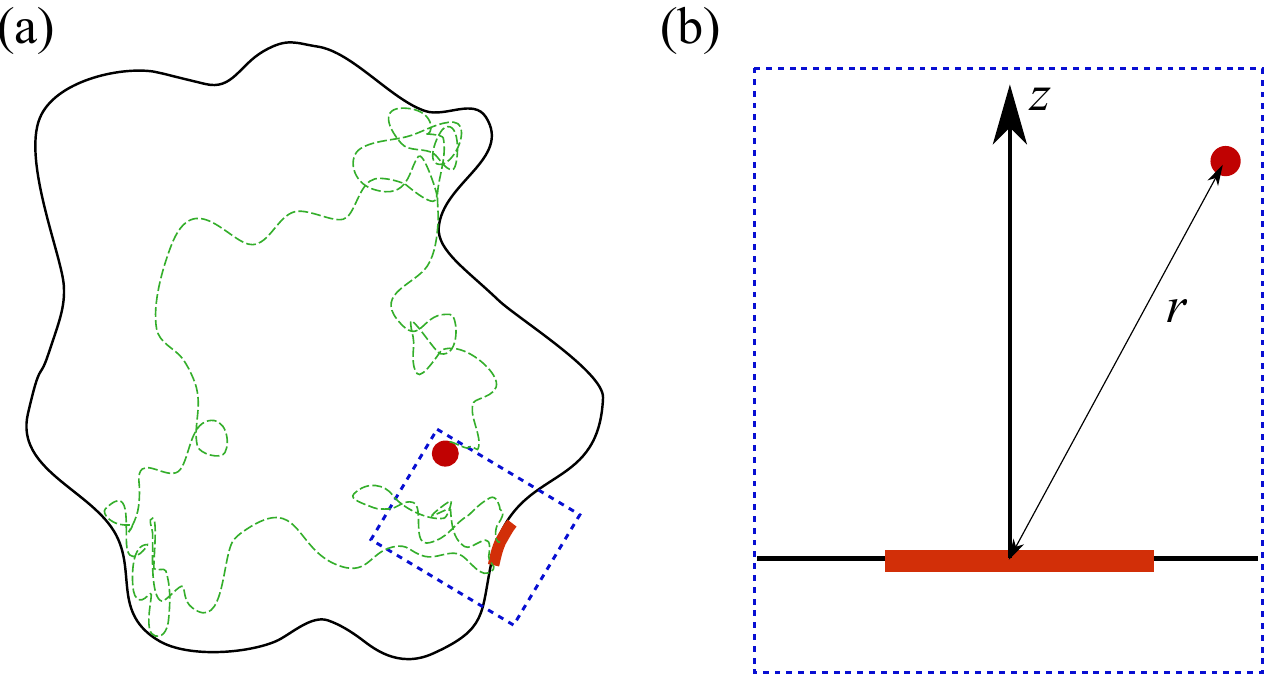}   
\caption{ (a) Illustration of the imperfect narrow escape problem. A partially reactive patch (thick red line) is embedded  in the boundary of a confining domain. A random walker, starting from the initial position (red sphere) diffuses in the domain and is eventually adsorbed on the patch. (b) Zoom on the portion of space delimited by the dashed blue lines around the reactive patch. }
\label{Fig1}
\end{figure}

Imperfect reactivity was early investigated for molecules diffusing in infinite space \cite{collins1949diffusion,traytak2007exact,Berg1985,shoup1981diffusion} (with an imposed concentration at infinity). The search problem for a single random walker moving in a \textit{confining} volume for an imperfect target, initially considered in Ref.~\cite{Szabo1980} for centered spheres, has  also attracted recent attention and several asymptotic results for imperfect search kinetics have been derived ~\cite{isaacson2016uniform,isaacson2013uniform,lindsay2017first,grebenkov2022statistics,grebenkov2022mean,chaigneau2022first,grebenkov2018strong,bressloff2022narrow,lindsay2015narrow}. Recently, explicit asymptotics of the   reaction time statistics have been obtained for  general Markovian random walks  \cite{guerin2021universal}.  
Besides the case of reactive targets located in the bulk of a confining domain, the \textit{narrow escape problem} (NEP)  consists in calculating the escape time of a random walker out of a confining domain, through a small window at the boundary of the domain (see Figure \ref{Fig1}(a)). While the NEP  is now well characterized for perfect reactions, for spherical domains \cite{Schuss2007,Singer2006a,Singer2006,singer2006narrow,mangeat2021narrow}  and large domains of arbitrary shapes \cite{Benichou2008}, fewer results are available for  imperfect reactions (i.e. for a partially adsorbing patch). The imperfect narrow escape problem has been investigated for particular geometries in  cylindrical~\cite{grebenkov2018towards,grebenkov2017effects} or spherical domains~\cite{grebenkov2017diffusive}\cite{grebenkov2019full} in which case the analysis depends on the eigenfunctions of the particular confining volume that is considered and relies on the so-called uniform flux approximation introduced in Ref.~\cite{shoup1981diffusion}. 

The aim of the present paper is to apply the formalism introduced in Ref.~\cite{guerin2021universal} to cover the case of the imperfect narrow escape problem in a domain of generic shape. Our formalism is asymptotically exact in the limit of large confining volume  -- it does not involve the constant flux approximation -- and provides explicit results in both regimes of small and large reactivity.  Of particular interest for imperfect reaction problems is the mean reaction time when the initial position is located on the reactive patch; this time is exactly zero for perfect reactions and scales as $1/\kappa$ for targets in the interior of the volume. We identify  a region for which the reaction time behaves anomalously with the reactivity $\kappa$. This region, which does not exist for targets in the bulk of the confining domain, is located at the boundary of the imperfectly reactive patch. While one would naively expect this time to be inversely proportional to $\kappa$, we find instead that when the initial position is at the \textit{boundary} of the reactive domain, the mean reaction time $\langle T\rangle_e$ is actually anomalously high ($\propto \kappa^{-1/2}$) and follows the exact asymptotics
\begin{align}
\langle T\rangle_e  \underset{\kappa\to\infty}{\sim}  \frac{V}{(d-1)   (2\pi \ \kappa D a)^{1/2} a^{d-2}} \label{Tbord}.
\end{align}
Here, $d=2$ or $d=3$ is the spatial dimension,  $D$ is the diffusion coefficient, $a$ is the radius of the reactive patch, and $V$ the volume of the confining domain. \mx{Here, we assume that the confining volume   is taken large enough, and the patch small enough, so that the confining boundaries at the vicinity of the target can be considered as a flat wall in which the reactive patch is embedded, the latter being considered as a line segment of length $2a$ in $d=2$ or as a flat disk of radius $a$ in $d=3$. } We show below how to obtain this anomalous scaling relation by solving a Wiener-Hopf integral equation. \mx{We will also show   how this ``anomalous'' behavior  (\ref{Tbord}) of $\langle T\rangle $ with $\kappa$ can be related to the divergence of fields in Laplacian problems near surfaces presenting asperities, as occurs in electrostatics near conducting edges \cite{jackson1977classical} or in the coffee ring effect \cite{deegan1997capillary}. }

More generally, we show that  the mean escape time for an arbitrary initial position far from the target, and for any finite reactivity $\kappa$ satisfies the following exact asymptotics:
\begin{align}
\langle T\rangle /V \underset{r\gg a}{\sim}
\begin{cases}
\frac{1}{ \pi D}\ln (r/a) + C_\infty & (d=2)\\
-\frac{1}{2\pi D r } + C_\infty & (d=3)
\end{cases},
\end{align}
where $C_\infty$ is independent of the initial distance $r$ from  the  target. For finite values of the reactivity $\kappa$, we show that $C_\infty$ can be obtained through a  semi analytical procedure. In the limit of large reactivity, we show that $C_\infty$ can be determined explicitly and is given by :
\begin{align}
C_\infty \underset{\kappa a\gg D}{\sim} 
\begin{cases}
\frac{\ln2}{\pi D}+ \frac{1}{\pi^2 \kappa a }\left( \ln\frac{8 \kappa a}{D}  + \gamma_e +1\right) & (d=2)\\
\frac{1}{4Da}+\frac{1}{4\pi \kappa a^2}\left( \ln\frac{2\kappa a}{D} +\gamma_e +1\right)     & (d=3)
\end{cases}
\label{BehavCInfty}
\end{align}
where $\gamma_e$ is Euler's constant. This expression is understood as the first two terms in the expansion of $C_\infty$ in powers of $1/\kappa$. Interestingly, this result shows that the term $C_\infty$ is not analytic in powers of $\kappa$, which originates from the anomalous scaling \eqref{Tbord}. Finally, we  also give exact results in the small reactivity limit, which will be found to be exactly the same (at first order) as the results obtained within the self-consistent, ``constant flux approximation'' that has been invoked to study the imperfect narrow escape in the literature~\cite{shoup1981diffusion,grebenkov2017diffusive,grebenkov2019full,grebenkov2018towards}. It is found that, far from the reactive patch this approximation is very accurate (for any reactivity), while it fails for initial positions  close to the reactive patch, and in particular  does not predict the behavior (\ref{Tbord}) near the boundary of the patch. 

The outline of the paper is as follows. First, we recall the formalism of Ref. \cite{guerin2021universal} in the particular case of the imperfect narrow escape problem for diffusing particles to obtain equations for the mean escape time in the large volume limit (Section \ref{LargeVolumeLimit}). 
We show how the formalism can be  presented under the form of an integral equation that is suitable for studying the large and small reactivity limits in Section \ref{SectionIntEq}. The large reactivity limit is investigated in Section \ref{SectionLargeKappa} where  Eqs.~(\ref{Tbord}) and (\ref{BehavCInfty}) are derived. 
\mx{In this Section, we also give a simple scaling argument that relates the anomalous behavior of $\langle T\rangle$ near the extremity of the patch to the divergence of electric fields near the edges of conducting objects. } The small reactivity limit is examined in Section  \ref{SmallKappaSect}.  
Last, we study briefly how the constant flux approximation can be implemented within our formalism in Section \ref{CFASection}. 
An exact, but formal solution for any reactivity parameter (that requires numerical tools, however) is presented in Appendix \ref{AppendixGeneralFormSolution}.


 

\section{Formalism for the imperfect narrow escape problem in the large volume limit}
\label{Formalism}

\subsection{General formalism}
\label{LargeVolumeLimit}

\begin{figure}[h!]
\includegraphics[width=7.5cm,clip]{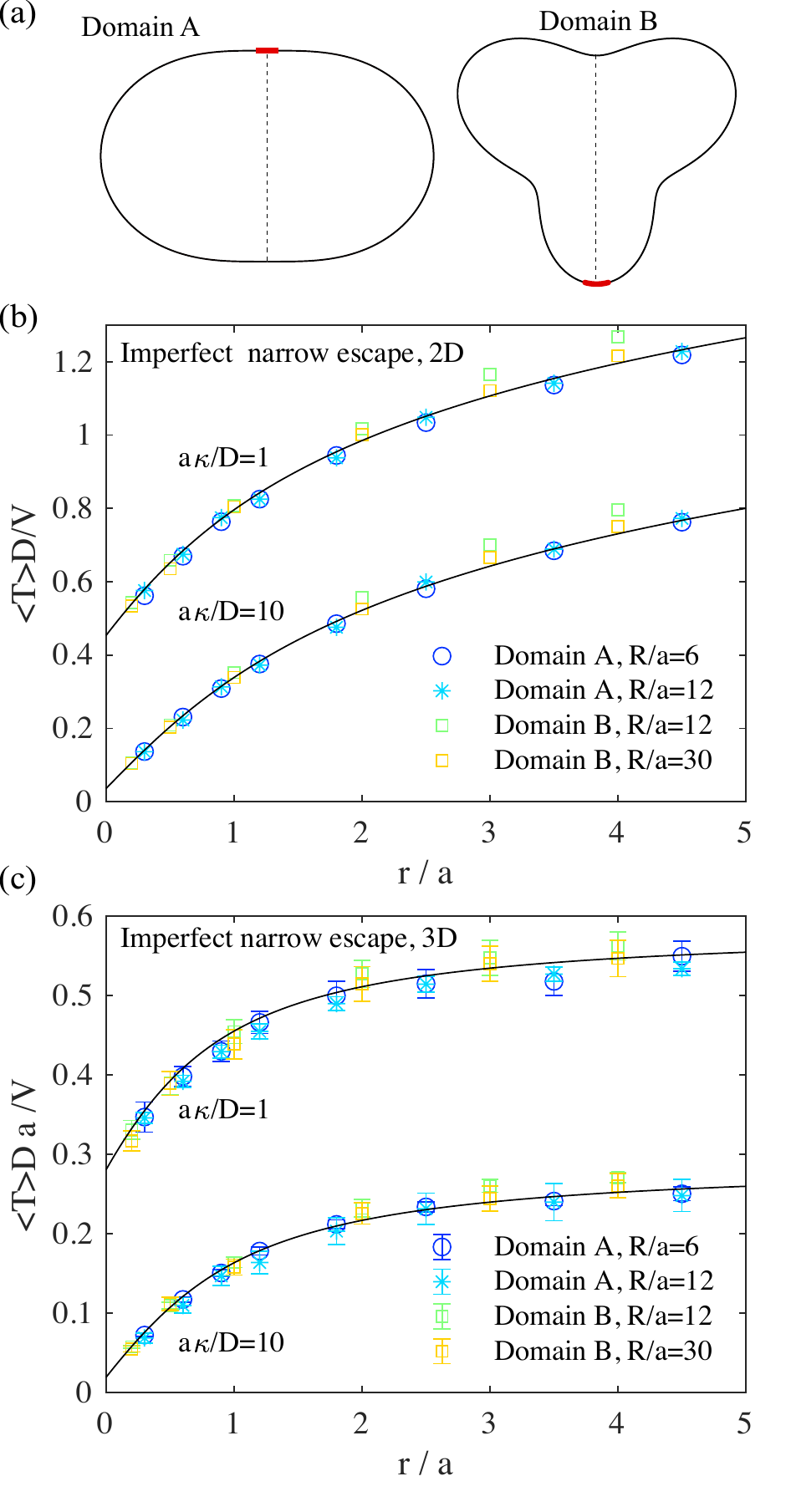}   
\caption{ (a) Geometry of the confining domains (called $A$ and $B$) that are considered for stochastic simulations. In 2D, these domains are defined in polar coordinates by $r(\theta)=R f(\theta)$ with $f=1.6(1+0.5\cos^2\theta)$ for domain $A$ and $f=1.6(1+0.1\sin\theta +0.3\sin3\theta)$ for domain $B$. Domains in 3D are obtained by considering revolution of 2D curves around the vertical dashed line. The reactive patch is indicated by a thick red line, and the initial position is taken at a distance $r$ from the center of the patch along the black dashed line. In the figure, we have used $R=6a$. (b),(c) Results of Brownian dynamics simulations for the mean reaction time in 2D/3D (parameters are indicated in the legend) compared to general theoretical expressions as obtained in Appendix \ref{AppendixGeneralFormSolution}. In all simulations, we used a time step $\Delta t=10^{-4}a^2/D$. Boundary conditions are implemented as follows: if, at the end of a time step, the random walker falls outside the domain, then if it is ``behind'' a reflecting wall it is reflected with respect to this wall, and if it falls ``behind'' the reactive patch, it is absorbed with probability $P_a=\kappa\sqrt{\pi dt/D}$ (in which case the trajectory ends) and it is reflected with probability $1-P_a$, see Ref.~\cite{singer2008partially}.}
\label{FigSimusNET}
\end{figure}

We consider the stochastic motion of an overdamped particle moving with diffusion coefficient $D$ in a confining volume $\Omega$. The boundary of the volume is $\partial\Omega$ and contains a small window $S_r$ which is partially reactive, the rest of the confining boundary is assumed to be smooth and perfectly reflecting, see Fig.~\ref{Fig1}(a).  \mx{We assume that $S_r$ is formed by the region of the surface at geodesic distance less than $a$ from the center, and $a$ is called the radius of the patch $S_r$.}
 The Fokker-Planck equation for the probability density $p(\ve[r],t)$ to observe the particle at position $\ve[r]$ and time $t$ is 
\begin{align}
&\partial_t p =D\ \nabla^2 p& (\ve[r]\in \Omega), \\
&\ve[n]\cdot \nabla p = 0  &(\ve[r]\in \partial\Omega \backslash S_r ),\\
& D\ \ve[n]\cdot \nabla p +\kappa \ p =0 & (\ve[r]\in S_r).
\end{align}
where $\ve[n]$ is the unit vector normal to the surface, pointing to the exterior of the volume. For a partially reactive surface, the reactivity parameter $\kappa$ is defined in such a way that the probability that the particle is absorbed by an infinitesimal surface element $dS$ located around $\ve[r]_s$ during $dt$ is $\kappa\ p(\ve[r]_s,t) dS dt$. 
We assume that the space dimension is $d=2$ (2D) or $d=3$ (3D). It is very well known that an equation for the mean first passage time can be obtained by identifying the adjoint transport operator \cite{Redner:2001a}, which in our case leads to the following equation for the mean reaction   time $\langle T\rangle(\ve[r])$ to the target, where $\ve[r]$ now represents the initial position of the particle:
\begin{align}
&D \nabla^2 \langle T\rangle =-1 & (\ve[r]\in\Omega)\label{EqMFPT},\\ 
& \ve[n]\cdot \nabla \langle T\rangle = 0  & (\ve[r]\in \partial\Omega \backslash S_r ),\\
& D\ \ve[n]\cdot \nabla \langle T\rangle +\kappa \langle T\rangle =0 & (\ve[r]\in S_r).
\end{align}
Integrating Eq.~(\ref{EqMFPT}) over the whole volume, and using the divergence formula and the boundary conditions leads to the exact integral relation: 
\begin{align}
\kappa \int_{S_r} dS(\ve[r]) \langle T\rangle = V, 
\end{align}
where $V=\vert\Omega\vert$ is the volume of the domain. 
Now, we consider the large volume limit, \mx{which is obtained when the confining volume extends without changing its shape, keeping constant the size of the target and the initial distance to the target.} We define the rescaled mean escape time $\Phi$ by
\begin{align}
\Phi(\ve[r] ) = \lim_{V\to\infty}\langle T(\ve[r])\rangle /V\label{ansatzPhi}.
\end{align}
In the large volume limit, the boundary at the vicinity of the reactive target becomes increasingly similar to a flat surface in which the reactive patch is a flat disk of radius $a$ in 3D (or a flat segment of length $2a$ in 2D).  Here, we denote the distance to the \mx{reflecting surface containing reactive patch} as $z$, see Fig.~\ref{Fig1}(b). With this in mind, inserting the ansatz (\ref{ansatzPhi}) into the above equations yields a closed system of equations in the large volume limit:
\begin{align}
&\nabla^2\Phi=0 \ \ (\text{if } z>0),  \label{EqPhi}\\
&\kappa \int_{S_r} dS(\ve[r])  \Phi=1, \label{NormCond} \\
&D\partial_z\Phi= \begin{cases}
0  & (\text{if } \ z=0,\vert\ve[r]\vert>a),\\
\kappa \ \Phi  & (\text{if } \ z=0,\vert\ve[r]\vert<a).\label{EqPhiB}
\end{cases}
 \end{align}
Importantly, we see that in the large volume limit, Eq.~\eqref{ansatzPhi}, the obtained equations are independent of the shape of the confining volume, which is present only though the scale factor $V$ in the definition of $\Phi$. We have directly controlled this aspect by performing numerical stochastic simulations of trajectories in the confined domain. The results of such simulations are shown on Fig.~\ref{FigSimusNET} and confirm that our formalism correctly predicts the mean reaction time in the large volume limit, independently on the shape of the confining domain. 

Equations (\ref{EqPhi}), (\ref{NormCond}), (\ref{EqPhiB}) generalize the formalism of Ref.~\cite{Benichou2008} to the case of imperfect reactions. The fact that the mean first reaction time scales with the volume is actually more general than the specific diffusive walk that we have considered here \cite{guerin2021universal}. To solve the above equations, we may be tempted to use spheroidal coordinates, which can be used to solve the problem for either infinite or vanishing reactivity. For finite reactivity however, the resulting equations in such coordinates involve Robin conditions with non-uniform coefficients, so that the mean reaction time can be obtained only in terms of the solution of an infinite linear system. 
This procedure is described in Appendix \ref{AppendixGeneralFormSolution}, and it indeed leads to a generic numerical solution that will be useful to test our analytical insights in all the paper. However, it is not suitable for analytical calculations. For this reason, we adopt a different approach, consisting in deriving an integral equation satisfied by $\Phi$ on the reactive patch.

\subsection{Obtaining an integral equation for the mean reaction time}
\label{SectionIntEq}
Let us first characterize the large distance behavior of the rescaled mean first reaction time $\Phi$. The condition (\ref{NormCond}), combined with the boundary condition at $z=0$, implies that 
\begin{align}
\int_{S_0} dS\ \ve[n]\cdot\nabla \Phi=1/D, \label{NormAlter}
\end{align}
for any surface $S_0$ whose intercept with the plane $z=0$ encloses the reactive patch. Taking such surface $S_0$ to be a half-disk of radius $R$ (in 2D) or a half-sphere (in 3D), we see that 
\begin{align}
\partial_r\Phi\underset{r\to\infty}{\sim}\begin{cases} 
1/[\pi D r] &(d=2),\\
1/[2\pi D r^2] & (d=3),
\end{cases}
\end{align}
where $r$ is the distance to the center of the reactive patch. Hence, the behavior of $\Phi$ for large $r$ takes the form 
\begin{align}
\Phi(\ve[r])\underset{\vert \ve[r]\vert\to\infty}{\sim}
\begin{cases}
\frac{1}{\pi D}\ln \vert \ve[r]\vert +C_\infty +o(1)  & (d=2), \\
-\frac{1}{2\pi D r } + C_\infty +o(1) & (d=3),
\end{cases} \label{LargeDistPhi}
\end{align}
where $C_\infty$ does not depend on $\ve[r]$.  The quantity $C_\infty$ thus characterizes the behavior of the mean reaction time far from the target, it could be used in matched asymptotics expansions if one aims to identify the first passage times distributions, as in ref. \cite{isaacson2016uniform} (which deals with interior targets). 

Let us now introduce the Green's function $G_N$ for the Laplace problem with \mx{Neumann} boundary conditions at $z=0$  (including on the reactive region). Such Green's function satisfy 
\begin{align}
&\nabla^2 G_N(\ve[r]\vert\ve[r]_0)=-\delta(\ve[r]-\ve[r]_0), \label{DefGN}\\
& \ve[n]\cdot\nabla G_N (\ve[r]\vert\ve[r]_0)=0 \hspace{2cm}  (z=0).\label{BC_GN}
 \end{align}
The expression of $G_N$ is easily found by using the image method \cite{barton1989elements}:
\begin{align}
G_N(\ve[r]\vert\ve[r]_0)=\begin{cases}
- \frac{1}{2\pi}[\ln\vert \ve[r]-\ve[r]_0\vert +\ln\vert \ve[r]-\ve[r]_0^*\vert]  & (d=2),\\
  \frac{1}{4\pi}\left[\frac{1}{\vert \ve[r]-\ve[r]_0\vert}+ \frac{1}{\vert \ve[r]-\ve[r]_0^*\vert}  \right] & (d=3) ,
\end{cases}  \label{GreenFunction}
\end{align}
where $\ve[r]_0^*$ represents the symmetric image of $\ve[r]_0$ with respect to the plane $z=0$. We note that the large $r$ behavior of $G_N$ is 
\begin{align}
G_N(\ve[r]\vert\ve[r]_0)\underset{\vert \ve[r]\vert\to\infty}=\begin{cases}
-\frac{1}{\pi}\ln\vert\ve[r]\vert+o(1/r) & (d=2) ,\\
\frac{1}{2\pi r} +o(1/r^2) & (d=3),
\end{cases}\label{largeDistGreen}
  \end{align}

We now use manipulations that are standard in Green's function problems \cite{barton1989elements} to put the problem for $\Phi$ on the form of an integral equation.  Using the Eq.~(\ref{DefGN}) and  $\nabla^2\Phi=0$ we  see that the following equality holds:
\begin{align}
\Phi(\ve[r]_0)=\int_{z\ge0} d\ve[r] [G_N(\ve[r]\vert\ve[r]_0) \nabla_{\ve[r]}^2\Phi -\Phi (\ve[r])\nabla_{\ve[r]}^2 G_N(\ve[r]\vert\ve[r]_0)].
\end{align}
Using the divergence formula, we   obtain
\begin{align}
& \Phi(\ve[r]_0)= \nonumber\\
& \int_S dS(\ve[r])\ve[n]\cdot  [G_N(\ve[r]\vert\ve[r]_0)\nabla_{\ve[r]}\Phi(\ve[r])-\Phi(\ve[r])  \nabla_{\ve[r]}G_N(\ve[r]\vert\ve[r]_0) ], \label{89542}
\end{align}
where $S$ is any closed surface in the half-space $z\ge0$. Taking this surface to be a half-circle (or half-sphere in 3D) of radius $R$ joined with a segment of size $2R$ on the axis $z=0$, we see that in the limit $R\to\infty$
\begin{align}
\Phi(\ve[r]_0)=C_\infty- \frac{\kappa}{D} \int_{S_r} dS(\ve[r]_s)  \Phi(\ve[r]_s) G_N(\ve[r]_s\vert \ve[r]_0), \label{IntEqForPhiAnyD}
\end{align}
where we have used  Eqs.~(\ref{LargeDistPhi}),(\ref{largeDistGreen}),(\ref{BC_GN})  and (\ref{EqPhiB})  to simplify the integrals over the surfaces  in Eq.~(\ref{89542}). The above equation means that $\Phi(\ve[r])$ can be constructed for any position as soon as one knows its value on the reactive patch. Taking $\ve[r]_0$ to be on the reactive patch yields an integral equation for $\Phi_s$, defined to be the value of $\Phi$ on the reactive patch. Since the above equation involves an unknown constant $C_\infty$ it must be accompanied by a supplementary condition, which is provided by the relation (\ref{NormCond}). Let us finally write explicitly the integral equations for $\Phi$ for the 2D and the 3D cases. For $d=2$, we obtain:
 \begin{align}
&\Phi_s(x_0)=C_\infty+ \frac{\kappa}{D \pi}\int_{-a}^a dx\   \Phi_s(x)\  \ln \vert x-x_0\vert, \label{IntegralEquation}\\
&\int_{-a}^a dx \ \Phi_s(x)=1/\kappa  \label{NormCond2D},
\end{align}
In the case $d=3$, we note that $\Phi_s(\ve[r])$ depends only on the radial distance to the disk center, $\Phi_s(\ve[r])=\Phi_s(r)$. The kernel of the integral equation can be simplified after a few algebraic manipulations detailed in Appendix \ref{AppendixEqInt3D}, leading to
 \begin{align}
&\Phi_s(r_0)=C_\infty- \frac{\kappa}{D  } \int_0^a dr \frac{2\ r \ \Phi_s(r) }{ \pi (r+r_0)} K\left(\frac{2 \sqrt{r r_0}}{r+r_0}\right),   \label{IntEq3D}  \\
& 2\pi \int_0^a dr  \ r\ \Phi_s(r)=1/\kappa ,
\end{align}
 where $K(k)$ is the complete elliptic integral of the first kind, defined as $K(k)=\int_0^1 dt [(1-t^2)(1-k^2 t^2)]^{-1/2}$ with $k$ the Elliptic modulus (to be distinguished from the parameter $m=k^2$). These integral equations admit no known analytical solution in general. In the next sections, we focus on their asymptotic study. From now on, without loss of generality, we set the units of length and time so that $a=1$ and $D/a^2=1$. The remaining parameter $\kappa$ then represents $\kappa a /D$ in full units.   


\section{The limit of large reactivity}
\label{SectionLargeKappa}

\subsection{A scaling argument for the anomalous behavior of the mean reaction time for large reactivity. }
\mx{Here we present a brief scaling argument that leads to the anomalous scaling (\ref{Tbord}). In the case of perfect reactions $\kappa=\infty$, it is clear from Eqs.~(\ref{EqPhi}),(\ref{EqPhiB}),(\ref{NormAlter}) that $\Phi$ can be seen as the electrostatic potential generated by a charged conducting disk that is embedded in an insulating surface, with the prescription that $\Phi=0$ on this disk. In fact, using the image methods, it is easy to show that $\Phi$ is also the electrostatic potential in infinite space generated by an infinitely thin disk. It is well known \cite{jackson1977classical} that, for this electrostatic problem, the ``electric field'' $-\nabla\Phi$ diverges near the edge as $1/\rho^{1/2}$, where $\rho$ is the distance from the disk extremity. 
Therefore, at a distance $\rho\ll a $ from the edges, $\Phi\propto\rho^{1/2}$. In the case that $\kappa$ is large but finite, since the natural length scale associated to finite reactivity is $\ell^*=D/\kappa$ \cite{grebenkov2019imperfect}, we may therefore assume that the mean reaction time is comparable to the mean first passage time when the starting position is located at a distance $\ell^*$ from the reactive patch. In  this condition, with $\rho=\ell^*$ we obtain the anomalous scaling $\Phi\propto1/\sqrt{\kappa}$, as announced in Eq.~(\ref{Tbord}). In what follows, we show how to rigorously derive this scaling law, with the prefactor. }   

\subsection{2D case}
\subsubsection{Leading order}
Consider now the limit $\kappa\to\infty$ in the case $d=2$. Since this situation corresponds to a first passage problem, we know that the asymptotic value of $C_\infty$ does not depend on $\kappa$. The fact that $\Phi_s$ vanishes in the large $\kappa$ limit leads us to postulate \mx{in line with Eqs.~\eqref{IntegralEquation}-\eqref{NormCond2D}} that 
\begin{align}
\Phi_s(x_0)\underset{\kappa\to\infty}{\sim} \frac{1}{\kappa} \Phi_{1}(x_0)  , \hspace{1cm} C_\infty\underset{\kappa\to\infty}{\sim}C_1\label{LeadingOrderScaling},
\end{align}
where $C_1$ and $\Phi_1$ do not depend on $\kappa$. Inserting these ansatz into the integral equation (\ref{IntegralEquation}) and the normalisation condition (\ref{NormCond2D}), we obtain
 \begin{align}
&\int_{-1}^1 dx  \ \Phi_{1}(x) \ln \vert x-x_0\vert=-\pi C_1, \label{IntEqPhi1} \\
&\int_{-1}^1 dx  \ \Phi_{1}(x) =1.
\end{align}
The integral equation  (\ref{IntEqPhi1}) for $\Phi_1(x)$ is known as Carleman's equation and its analytical solution is known  explicitly \cite{HandbookIntegralEquations}. Using also the normalisation condition,
we obtain the final expression for $C_1$ and $\Phi_1$:
\begin{align}
C_1=\frac{\ln2}{\pi}, \ 
\Phi_{1}(x)= \frac{1}{\pi \sqrt{1-x^2} }  \label{PHIM1}.
\end{align}

\subsubsection{Boundary layer near the extremities of the reactive patch}
We now note that $\Phi_{1}(x)$ is formally infinite at $x=\pm1$, i.e. near the boundary of the reactive patch. This means that our expansion (\ref{LeadingOrderScaling}) is not valid near these points, suggesting a behavior similar to those obtained for boundary layer problems. Since the reaction length \cite{grebenkov2019imperfect} is $1/\kappa$ in our units, we  expect processes happening at such scales. Therefore, we   assume the behavior 
\begin{align}
\Phi_s(x)=\kappa^\alpha \psi( (1-\vert x\vert)\kappa), \hspace{1cm} (1-\vert x\vert) \ll1 \label{ExpBL},
\end{align}
with $\psi$ a scaling function. The exponent $\alpha$ will be set such that the behavior of $\Phi$ in the boundary layer matches with that far from the boundary layer. Namely, the compatibility of the above ansatz with Eq.~(\ref{PHIM1}) imposes the choice 
\begin{align}
\alpha=-1/2,\hspace{1cm}\psi(X)\underset{X\to\infty}{\sim}\frac{1}{\pi\sqrt{2X}}\label{MatchingCond2D}.
\end{align}
Hence, the structure of the solution in the limit $\kappa\to\infty$ is  
\begin{align}
\Phi_s(x)=\begin{cases}
\kappa^{-1}\Phi_{1}(x) +... & (1-\vert x\vert)\gg 1/\kappa\\
\kappa^{-1/2}\psi( (1-\vert x\vert)\kappa)+...  & (1-\vert x\vert)\ll 1 
\end{cases}\label{Struct2D}
\end{align}
and the condition (\ref{MatchingCond2D}) ensures that these two expressions give the same result in their common validity regime \mx{$\kappa^{-1}\ll 1-\vert x\vert\ll 1$}. 
A key point here is that the mean return time, starting from the boundary of the reactive region scales as $1/\kappa^{1/2}$ and is thus infinitely larger than the mean return time starting from the center of the target, which scales as $1/\kappa$.  
This suggests to set $C_\infty=C_1+C_{1/2}/\sqrt{\kappa}+...$ in the limit of large reactivity (even though the constant $C_{1/2}$ will turn out to vanish).  

Let us now find the set of equations satisfied by $\psi$. First, we consider the normalisation condition (\ref{NormCond2D}), which we write under the form
\begin{align}
\int_{-1}^1 dx \left[\Phi_s(x)-\frac{\Phi_{1}(x)}{\kappa}\right]=0\label{NormCond1}.
\end{align}
\mx{Here we remark that the integrand in the above integral is maximal near $x=\pm1$. Let us define an \textit{intermediate length scale} $\ell$ such that 
\begin{align}
1/\kappa\ll\ell\ll1\label{DefEll}.
\end{align}
We will keep this notation in the whole paper. We write Eq.~\eqref{NormCond1} by separating the integral into two regions:  when $(1-\vert x\vert)>\ell$ then we approximate $\Phi_s(x)$ by the first line of Eq.~(\ref{Struct2D}) and if $(1-\vert x\vert)<\ell$ we use the expressions on the second line of  Eq.~(\ref{Struct2D}) for $\Phi_s$ and we approximate by its behavior near $x=\pm1$, which reads $\Phi_{1}\simeq 1/[\pi\sqrt{2(1-\vert x\vert)}]$. This leads to
\begin{align}
0&= \int_{-1}^{-1+\ell} dx \left[\frac{\psi((1-\vert x\vert )\kappa)}{\sqrt{\kappa}}-  \frac{1}{\kappa\pi\sqrt{2(1-\vert x\vert)}}\right]\nonumber\\
&+ \int_{1-\ell}^{1} dx \left[\frac{\psi((1-\vert x\vert) \kappa)}{\sqrt{\kappa}}-\ \frac{1}{\kappa\pi\sqrt{2(1-\vert x\vert)}}\right].\nonumber
\end{align}
Setting $X= (1-\vert x\vert)\kappa$, and using $\ell\kappa\gg1$, we obtain}
\begin{align}
\int_{0}^\infty dX \left[\psi(X)-\frac{1}{\pi\sqrt{2X}}\right]=0 \label{Cond}.
\end{align}
In order to find the equation satisfied by $\psi$ it is useful to write the difference between the general equation (\ref{IntegralEquation}) and the equation satisfied by $\Phi_{1}$, to find
\begin{align}
\Phi_s(x_0)=&\ C_\infty-C_1 \nonumber \\
&+ \frac{\kappa}{\pi}\int_{-1}^1 dx \left[\Phi_s(x)-\frac{\Phi_{1}(x)}{\kappa}\right] \ln\vert x-x_0\vert. \label{StartingPointIntEq}
\end{align}
 This leads to
\begin{align}
&\Phi_s(x_0)= \ C_\infty-C_1 \nonumber\\
&+ \frac{\kappa}{\pi}\int_{-1}^{-1+\ell} dx \left[\frac{\psi((1-\vert x\vert )\kappa)}{\sqrt{\kappa}}-\frac{\Phi_{1}(x)}{\kappa}\right] \ln\vert x-x_0 \vert \nonumber\\
&+ \frac{\kappa}{\pi }\int_{1-\ell}^{1} dx \left[\frac{\psi((1-\vert x\vert) \kappa)}{\sqrt{\kappa}}-\frac{\Phi_{1}(x)}{\kappa}\right] \ln\vert x-x_0\vert.
\end{align}
For $x_0=1- X_0/\kappa$, if we set $X= (1-\vert x\vert)\kappa$, this yields 
\begin{align}
& \psi(X_0) =C_{1/2}
+ \int_0^{\ell\kappa} \frac{dX}{\pi}  \left[\psi(X)-\frac{1}{\pi\sqrt{2X}}\right]\ln  \frac{\vert X-X_0 \vert}{\kappa}    \nonumber\\
&+ \frac{1}{\pi}\int_0^{\ell\kappa} dX \left[\psi(X)-\frac{1}{\pi\sqrt{2X}}\right]   \ln\left\vert 2-\frac{X+X_0}{\kappa}\right\vert  .
\end{align}
Now, in the limit $\kappa\to\infty$, noting that $\ell\kappa\to\infty$ [by definition of $\ell$, see Eq.~(\ref{DefEll})] and using the previously found condition (\ref{Cond}) we obtain 
\begin{align}
\psi(X_0)&=C_{1/2}  \nonumber\\
&+\int_0^{\infty} dX \left[\psi(X)-\frac{1}{\pi\sqrt{2X}}\right] \frac{ \ln   \vert X-X_0\vert}{\pi}.  \label{WHEq}
\end{align}
This is a Wiener-Hopf integral equation for the unknown function $\psi(X)$. We solve it in the next section. 

\subsubsection{Solution of the Wiener-Hopf equation (\ref{WHEq})}
\begin{figure}[h!]
\includegraphics[width=8cm,clip]{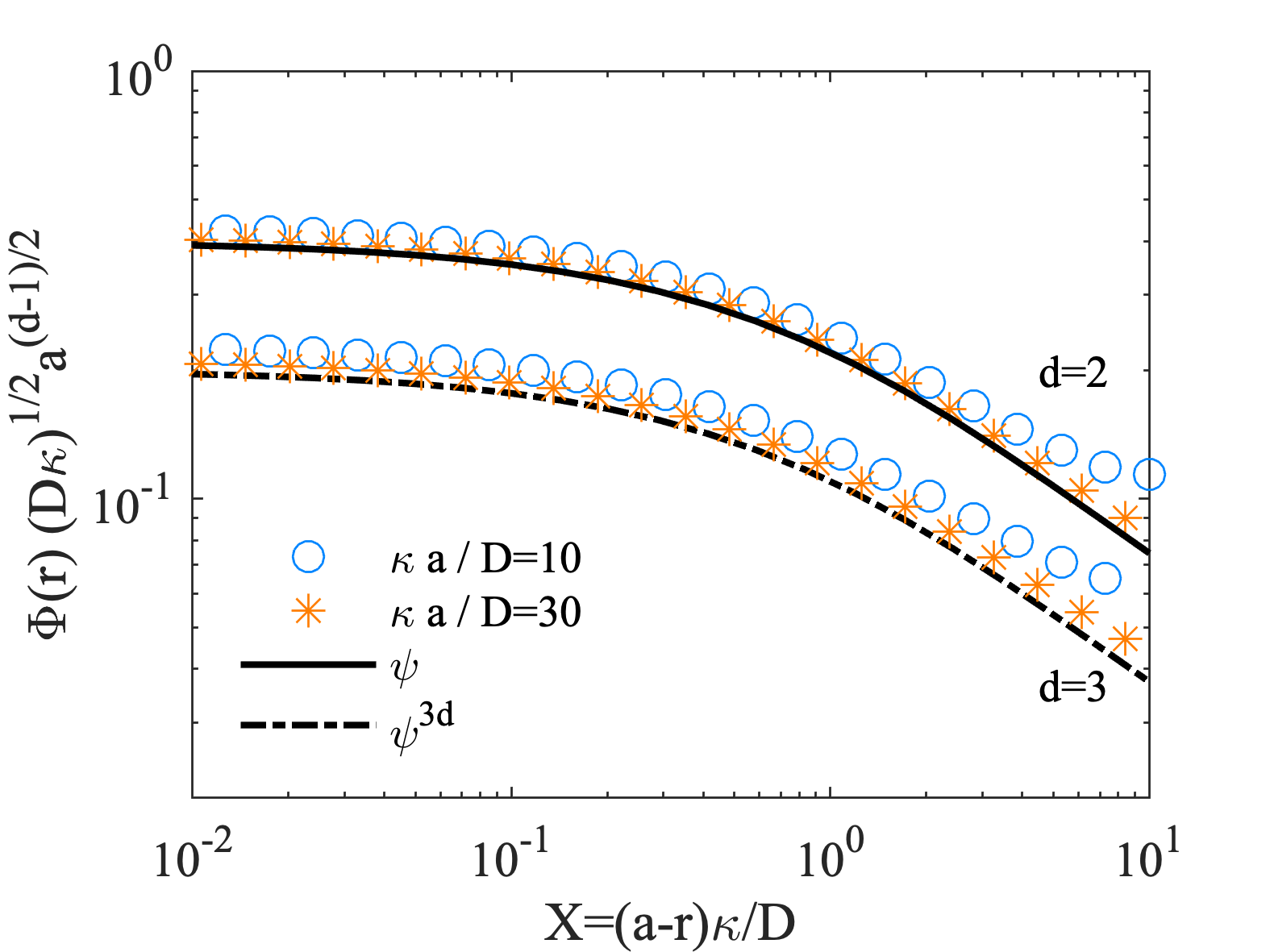}   
\caption{Behavior of the mean first reaction time near the extremities of the reactive patch. Symbols: exact general solution obtained numerically in Appendix \ref{AppendixGeneralFormSolution} in 2D (upper symbols) and 3D (lower symbols). We also represent the values of $\psi$ and $\psi^{3d}$ obtained from Eqs.~(\ref{ExplicitCalculationPsi2D}) and (\ref{psi3D}). }
\label{Fig_NET_CheckPsi}
\end{figure}

We   solve the  Wiener-Hopf integral equation with Carleman's method, as described in Ref.~\cite{HandbookIntegralEquations}. We note that a similar equation has appeared in viscous flow theory \cite{brown1977,boersma1978,kida1994} but the differences between our equations and the equation studied in these references justify the fact to solve it here in detail. 
First, let us introduce the following notations. We \mx{denote} $f_+(X)$ all functions (\mx{depending on} the real variable $X$) that vanish for all $X<0$, and $f_-(X)$ all functions that vanish for $X>0$. For any function $f(X)$ one can write $f(X)=f_+(X)+f_-(X)$, with $f_+(X)=f(X)\theta(X)$ and $f_-(X)=f(X)\theta(-X)$, where $\theta$ is the Heaviside step function. We introduce the complex Fourier transform and its inverse:
\begin{align}
&\hat{f}(z)=\int_{-\infty}^\infty dX f(X) e^{-i z X}, \\ 
&f(X)=\frac{1}{2\pi}\int_{-\infty}^\infty du \hat{f}(u) e^{+i u X} 
\end{align}  
where $z$ represents a complex number and $u$ a real number. 
We denote $\hat{f}_+(z)$ the Fourier transform of the function $f_+(X)$, and  $\hat{f}_-(z)$ the Fourier transform of $f_-(z)$. Typically, Fourier transforms of the form $\hat{f}_+(z)$ are defined in the lower complex half-plane $\text{Im}(z)\le 0$,   Fourier transforms of the form $\hat{f}_-(z)$ are defined in the upper complex half-plane $\text{Im}(z)\ge 0$ (as long as $f_{\pm}(x)$ does not diverge exponentially at $x\to\pm\infty$). Now, we can define $\psi_+(X)\equiv\psi(X)\theta(X)$, and we introduce
\begin{align}
K(X)=\frac{1}{\pi}\ln \vert X\vert, \hspace{1cm}
\chi_+(X)=\frac{1}{\pi\sqrt{2X}}\theta(X). 
\end{align}
The integral equation (\ref{WHEq}) can be generalized for negative $X_0$ by writing
\begin{align}
&\psi_+(X_0)=    \nonumber\\
&\int_0^{\infty} dX \left[\psi_+(X)-\chi_+(X)\right]  K(X-X_0)    + y_-(X_0), 
\end{align}
where the only remarkable property of $y_-(X_0)$ is that it vanishes for positive $X_0$. Note that we have assumed that $C_{1/2}=0$, this will be justified at the end of the calculation by the fact that the obtained solution satisfies the normalization condition (\ref{Cond}) for this value of $C_{1/2}$.
Taking the Fourier transform of the above equation, we obtain 
\begin{align}
\hat{\psi}_+(u)=    [\hat{\psi}_+(u)-\hat{\chi}_+(u) ]  \hat{K}(u)   + \hat{y}_-(u).
\end{align}
Calculating the Fourier transforms leads to
\begin{align}
\hat{\psi}_+(u)\left(1+\frac{1}{\vert u\vert}\right) = -  \frac{1-i \ \text{sign}(u)}{2 \sqrt{\pi } | u |^{3/2}}    + \hat{y}_-(u) ,
\end{align}
where $\text{sign}(u)=\theta(u)-\theta(-u)$ is the sign function.  
This equation can be considerably simplified by introducing an auxilliary function $S_-(X)$ defined by
\begin{align}
S_-(X)=\theta(-X) \frac{  \sqrt{2 \left| X\right| }}{\pi } ,\ \hat{S}_-(u)=\frac{1 -i \ \text{sign}(u)}{2 \sqrt{\pi } \left| u\right| ^{3/2}},
\end{align}
so that the Wiener-Hopf equation can be written as 
\begin{align}
\hat{\psi}_+(u)\left(1+\frac{1}{\vert u\vert}\right)  =  - \hat{S}_-(u) + \hat{y}_-(u) = \hat{f}_-(u),
\end{align}
where $\hat{f}_-(u)$ is the Fourier transform of another unknown function $f_-(X)$, whose only remarkable property is to vanish for positive $X$. This kind of equations is known as a homogeneous Wiener-Hopf equation, the method to solve it consists in obtaining a factorization of the form $\hat{\psi}_+(u)\hat{g}_+(u)=\hat{g}_-(u)$. To this end, we  write the Wiener-Hopf equation under the form
\begin{align}
\hat{\psi}_+(u) e^{\hat{W}(u)} =\hat{f}_-(u)  \label{9421},
\end{align}
with
\begin{align}
\hat{W}(u)=\ln [1+1/\vert u\vert] ,
\end{align}
and we seek a factorization $\hat{W}(u)=\hat{W}_+(u)+\hat{W}_-(u)$. This can be done by calculating its inverse Fourier transform:
\begin{align}
W(X)=\frac{ \cos (X) [2\ \text{Si}(\left| X\right| )-\pi]+\pi }{2 \pi  \left| X\right| } 
-\frac{\text{Ci}(\left| X\right| ) \sin ( X ) }{ \pi    X  }
\end{align}
 where Ci and Si are the integral cosine and integral sine functions
\begin{align}
\text{Ci}(X)=-\int_X^\infty dt \frac{\cos(t)}{t}, \hspace{0.3cm} \text{Si}(X)=\int_0^X dt \frac{\sin(t)}{t}.
\end{align}
A factorization may thus be obtained by setting $W(x)=W_+(x)+W_-(x)$, i.e $W_+(x)=W(x)\theta(x)$ and $W_-(x)=W(x)\theta(-x)$. 
Now, we write the equation (\ref{9421}) as
\begin{align}
\hat{\psi}_+(u) e^{\hat{W}_+(u)} =\hat{f}_-(u)e^{-\hat{W}_-(u)}. 
\end{align}
We are now in the favorable case: the terms on left-hand-side are analytic functions in the upper complex plane, those on the right are analytic in the lower complex plane except for one pole at $z=0$, and these terms are equal on the real axis. According to the theorem of analytic continuation, combined with the Cauchy theorem, we conclude that both terms are equal to a constant plus a $1/z$ term on the whole complex plane \cite{HandbookIntegralEquations}. We thus have 
\begin{align}
\hat{\psi}_+(u) e^{\hat{W}_+(u)} =a_0 +\frac{a_1}{u},
\end{align}
where the constants $a_0,a_1$ will be found by requiring that $\psi(X)$ is a solution to our problem. Since $\hat{W}_+(z)$ is defined on the lower complex plane, we may consider the above equations on the lower imaginary axis $u=-i s$, in which case the above equality can be written in terms of Laplace transforms, with the usual notation $\tilde{f}_+(s)=\int_0^\infty dt e^{-st}f_+(t)=\hat{f}_+(-i s)$:
\begin{align}
\tilde{\psi}_+(s)  =\left(a_0 +\frac{i\ a_1}{s}\right) e^{-\tilde{W}^+(s)} \label{954321}.
\end{align}
The Laplace transform $ \tilde{W}^+(s)$ can be identified by calculating its derivative, i.e. the Laplace transform of $-x W(x)$, and then by integrating over $s$; this leads to
\begin{align}
\tilde{W}^+(s)=\frac{1}{4}\ln\frac{1+s^2}{s^2}+m(s) \label{EqWPlus},
\end{align}
with 
\begin{align}
m(s)=-\int_0^s \frac{dw \ \ln w}{\pi(1+w^2)}.
\end{align}
We know that the behavior of $\psi$ for large arguments is given by the matching condition Eq. (\ref{MatchingCond2D}), which translates to the small-$s$ behavior:
\begin{align}
\tilde{\psi}_+(s)\underset{s\to0}{\sim}\frac{1}{\sqrt{2 \pi s}} \label{SMallsPsi} .
\end{align}
Inserting Eq. (\ref{EqWPlus}) into  (\ref{954321}) and taking the small-$s$ limit, we see that the above behavior is  obtained for $ia_1=1/\sqrt{2\pi}$. Next, the value of $a_0$ is found by requiring that $\psi(0)$ is finite, so that $\tilde{\psi}(s)$ vanishes in the limit $s\to\infty$; this leads to $a_0=0$.
 Hence, the final expression for the function $\psi$ is given in the Laplace domain by
\begin{align}
\tilde{\psi}_+(s)=\frac{1}{\sqrt{2\pi s}(1+s^2)^{1/4}}e^{-m(s)}  .
\end{align}
Finally, we must check that the normalization condition  (\ref{Cond}) holds, this is  readily done by noting that $\tilde{\psi}(s)-1/\sqrt{2\pi s}=\mathcal{O}(\sqrt{s}\ln s)$ vanishes for small $s$. This justifies our hypothesis $C_{1/2}=0$.  Unfortunately, the Laplace inversion cannot be performed so that we know only $\psi(X)$ in closed form, however we can easily derive the asymptotic behavior of $\psi(X)$ for small and large arguments. 
The study of the asymptotic behavior of $\tilde{\psi}(s)$ for large $s$ leads readily to the initial value of $\psi(X)$:
\begin{equation}
\psi(0)=\lim_{s\to\infty}s\tilde{\psi}(s)=\frac{1}{\sqrt{2\pi}}, 
\end{equation}
this justifies the previously announced result (\ref{Tbord}). The large $X$ behavior can be computed by noting that the Laplace transform of $X\psi(X)$ is $d\tilde{\psi}/ds$ and by expanding this one for small $s$, with the result:
\begin{align}
\psi(X)\underset{X\to\infty}{\simeq} \frac{1}{\sqrt{2X} \pi }+\frac{ \ln (4X)+\gamma_e -1}{ \pi ^2(2X)^{3/2}}+...\label{PsiLargeX}
\end{align}
with $\gamma_e$ the Euler-Mascheroni constant. 
A formula for $\psi(X)$ can be obtained by considering the inverse Laplace transform of $\tilde{\psi}(s)$ with the Mellin's inverse formula, by using a contour that follows the  negative real axis (above and below), such Laplace inversion is obtained  in Appendix 4 of Ref.~\cite{kida1994}:
\begin{align}
\psi(X)=\frac{1}{\pi } \int_0^{\infty} dp\  \frac{e^{- p X+m(p)}}{\sqrt{2\pi p }(1+p^2)^{3/4}   }   . \label{ExplicitCalculationPsi2D}
\end{align}
In summary, here we have obtained an analytic expression in Laplace space for the scaling function $\psi(X)$ which characterizes the behavior of the mean first passage time near the extremities of the reactive patch in two dimensions. The validity of our approach is checked on Figure \ref{Fig_NET_CheckPsi} by comparing with exact numerical results obtained from the general form of the solution.

\subsubsection{Next-to-leading order expansion}

Up to now, the constant $C_\infty$, which characterizes the behavior of the mean reaction time when the initial position is far from the target, has been obtained at leading order only in the large reactivity limit, with the same result as in the case of a perfectly reactive patch. Here we show how to obtain the first non-trivial correction for $C_\infty$ for large reactivity, with the result that  $C_\infty(\kappa)$ is not analytic in $\kappa$.
First, we note that when $1/\kappa\ll 1-\vert x\vert \ll 1$, Eqs.~(\ref{PsiLargeX}) and (\ref{Struct2D}) indicate that   
\begin{align}
\Phi_s(x)\simeq \frac{1}{\kappa\pi \sqrt{2 (1-\vert x\vert)}  }+\frac{ \mx{\ln [4 (1-\vert x\vert)\kappa]}+\gamma_e -1}{\kappa^2 \pi ^2[2( 1-\vert x\vert)]^{3/2}}+...\label{ITREZ9543}
\end{align}
 This suggests that, outside the boundary layer, the next-to-leading order behavior of $\Phi_s$ reads:
 \begin{align}
 \Phi_s(x)= \frac{\Phi_1}{\kappa}+\frac{\Phi_2^*\ln\kappa+\Phi_2}{\kappa^2}+... ,
 \end{align}
because this expression can be matched with (\ref{ITREZ9543}) by imposing that 
 \begin{align}
& \Phi_2^*( x)\simeq \frac{  1}{  \pi ^2[2  (1-\vert x\vert)]^{3/2}}, &( x\to\pm1),\label{MatchCond2Star} \\ 
 &\Phi_2( x)\simeq \frac{ \ln [4 (1-\vert x\vert )]+\gamma_e -1}{ \pi ^2[2 (  1- \vert x\vert)]^{3/2}},&( x\to\pm1).\label{MatchCond2}
 \end{align}
These expansions also lead us to assume that, for large reactivity the constant $C_\infty$ behaves as
 \begin{align}
& C_\infty=C_1+\frac{C_2+C_2^*\ln \kappa}{\kappa}+...& (\kappa\to\infty)
 \end{align}
 
\begin{figure}[h!]
\includegraphics[width=8cm,clip]{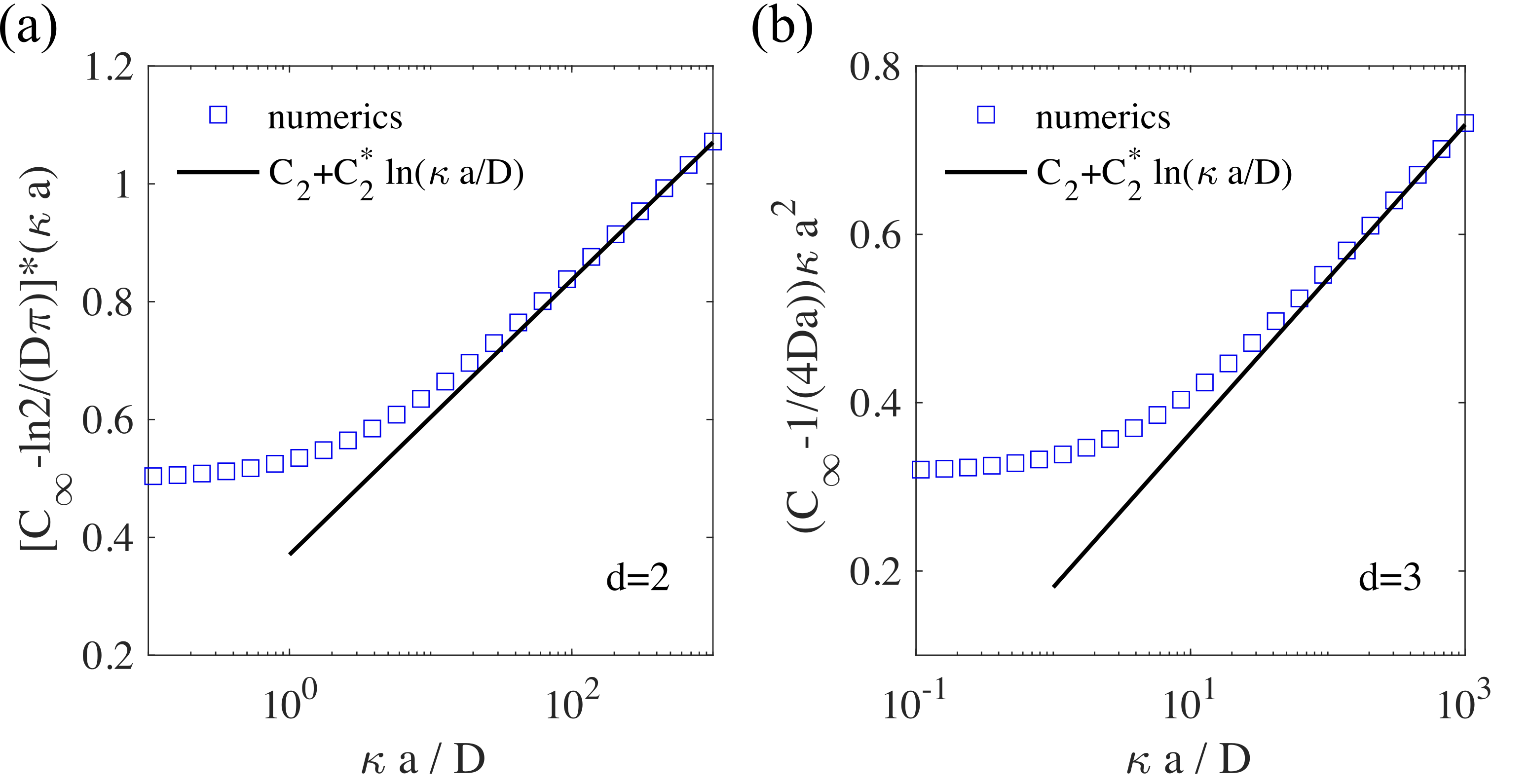}   
\caption{Comparison of the values of $C_\infty$ obtained numerically (symbols, see Appendix~\ref{AppendixGeneralFormSolution} for details),  with the analytical predictions in Eqs.~(\ref{EqC2_2D}) and (\ref{ValueC2_3D}) (black lines), for the two-dimensional (a) and three-dimensional (b) domains.}
\label{FigCheckC2andC2star}
\end{figure}

The equation for $\Phi_2$ and $\Phi_2^*$ can be identified as follows. We consider again the intermediate length scale $\ell$ satisfying (\ref{DefEll}), and we start from the integral equation (\ref{StartingPointIntEq}), which we write as
\begin{align}
 &\int_{-1+\ell}^{1-\ell} dx \left[\frac{\Phi_2^*(x)\ln \kappa +\Phi_2(x)}{\kappa}\right] \ln\vert x-x_0\vert  = \nonumber\\
&-\pi \frac{C_2+C_2^*\ln \kappa}{\kappa}+ \frac{\pi \phi_1(x_0)}{\kappa}+ B(x_0)+B(-x_0)  \label{TruncatedIntegralEq},
\end{align}
where $B$ contains all the terms which appear due to the fact that the integral over $\Phi_2$ in the above equation is evaluated over a truncated interval $]-1+\ell;1+\ell[$ instead of $]-1;1[$, so that
\begin{align}
B=& \int_0^{\ell\kappa} \frac{ dX}{\sqrt{\kappa} }  \left(\frac{1}{\pi\sqrt{2X}}-\psi(X)\right) \ln\left(   1 - x_0 -\frac{X}{\kappa} \right)   \label{DefB}.
\end{align}
To proceed further, we consider (\ref{TruncatedIntegralEq}) as an integral equation for $\Phi_2+\ln\kappa \Phi_2^*$ over the truncated interval  $]-1+\ell;1+\ell[$. Its solution is analytically known and we identify the constants $C_2$ and $C_2^*$ by requiring that the normalisation condition is satisfied at this order of $\kappa$. This procedure requires to evaluate $B$ in the limit $\ell\to0$ without assuming that $\ell\ll 1-\vert x_0\vert$, and it turns out to be relatively tedious. The calculation is described in Appendix \ref{CalculationC2_2D}, and leads to the explicit results 
\begin{align}
&C_2^*=\frac{1}{\pi^2}, &
C_2=\frac{1+\gamma_e+\ln 8}{\pi^2}.\label{EqC2_2D}
\end{align} 
These values of $C_2$ and $C_2^*$ are in excellent agreement with the exact solution for $\Phi_s$ obtained numerically, as shown in Fig.~\ref{FigCheckC2andC2star}(a). 

\subsection{3D case}  
We now adapt the approach to the 3D case. It turns out that the solution admits the same scaling behaviors than in 2D:
 \begin{align}
 \Phi_s(r)= \begin{cases}
 \frac{\Phi_1(r)}{\kappa}+\frac{\Phi_2^*(r) \ln\kappa+\Phi_2(r)}{\kappa^2} & (1-r) \gg 1/\kappa,  \\
\frac{1}{\sqrt{\kappa}}\psi^{3d}( (1-r) \kappa)    & (1- r) \ll 1,
 \end{cases} \label{Ansatz3D}
 \end{align}
 where the first line is the expansion of $\Phi_s(r)$ in powers of $\kappa$ at fixed $r$, and the second line the expansion of $\Phi$ in powers of $\kappa$ at fixed $X=(1-r)\sqrt{\kappa}$. At leading order, the integral equation for $\Phi_1$ reads
\begin{align}
&0=C_1 -   \int_0^1 dr\    \frac{2r}{ \pi (r+r_0)} K\left(\frac{2 \sqrt{r r_0}}{r+r_0}\right) \Phi_1(r),  \\
&2\pi \int_0^1 dr\ r\ \Phi_1(r) =1. 
\end{align}
The solution of the above integral equation (where $r\ \Phi_1(r)$ is considered to be the unknown function) is known \cite{HandbookIntegralEquations} and this leads to the solution
\begin{align}
\Phi_1(r)=\frac{1}{2 \pi  \sqrt{1-r^2}}, \hspace{1cm} C_1=\frac{1}{4}.\label{954230}
\end{align}
We thus note that 
\begin{align}
\Phi_1(r\to 1)\sim \frac{1}{2 \pi\sqrt{2(1- r )}}.\label{954231}
\end{align}
In the boundary layer near $r=1$, we set $r=1-X/\kappa$, $r_0=1-X_0/\kappa$, and $\Phi_s(r)=1/\sqrt{\kappa}\psi(X)$. With these scalings we can expand the integral equation  (\ref{IntEq3D}) with the result that    $\psi^{3d}$ satisfies exactly the same equation than in 2D, the only difference is that it has to match with $\psi^{3d}(X)\sim 1/(2\pi\sqrt{2X})$ for large $X$ [due to Eq.~(\ref{954231})] and there is thus a factor of 2 that arises when we compare to the 2D case:
\begin{align}
\psi^{3d}(X)=\frac{1}{2}\psi (X)\label{psi3D}.
\end{align} 
This  relation is checked on Fig.~\ref{Fig_NET_CheckPsi}.  
Let us now identify the next order terms in 3D. Inserting the ansatz (\ref{Ansatz3D}) into    the integral equation  (\ref{IntEq3D})  and expanding at second order, we obtain
\begin{align}
\int_0^{1-\ell}  \ \frac{dr\ r}{r+r_0}K\left(\frac{2\sqrt{r r_0}}{r+r_0} \right)[ \Phi_2(r)+\Phi_2^*(r)\ln\kappa]=\nonumber\\
-\frac{\pi\Phi_1}{2}+\frac{\pi}{2}[C_2^*\ln\kappa+C_2]+B\label{5843}
\end{align}
where the term $B$ compensates the fact that the above  integrals are evaluated  over the  truncated interval $[0;1-\ell[$, so that:
\begin{align}
B(r_0,\ell)= -    \sqrt{\kappa}\int_{1-\ell}^1  \ \frac{dr\ r}{r+r_0}K\left(\frac{2\sqrt{r r_0}}{r+r_0} \right) \nonumber\\
\times
\left[\psi^{3d}((1-r)\kappa)-\frac{1}{2\pi \sqrt{2(1-r)\kappa} }\right]\label{DefB3D}.
 \end{align}
As in the 2D situation, we consider (\ref{5843}) as an integral equation for which the solution is analytically known; and we then chose $C_2$ and $C_2^*$ so that the normalisation condition for $\Phi_s$ holds at all orders of $\kappa$. The final result is
\begin{align}
C_2^*=\frac{1}{4\pi}, \ 
C_2=\frac{\gamma_e +1+\ln (2)}{4 \pi }, \label{ValueC2_3D}
\end{align}
and it agrees perfectly with numerical solutions, as shown in Fig.~\ref{FigCheckC2andC2star}(b).

\section{The limit of small reactivity}
 \label{SmallKappaSect}
Let us now consider the limit $\kappa\to0$. At leading order, the mean reaction time is homogeneous. We seek a solution under the form
\begin{align}
\Phi_s(x)=\frac{1}{\kappa }\sum_{n\ge0} f_n(x)\kappa^n, \ C_\infty=\frac{1}{\kappa }\sum_{n\ge0} c_n \kappa^n.
\end{align} 
At leading order, we obtain
\begin{align}
f_0=c_0=1/\vert S_r\vert . 
\end{align}
where $S_r$ is the length (in 2D) or the area (in 3D) of the reactive patch. Furthermore, next-orders can be found iteratively by using
\begin{align}
f_n(\ve[r])=c_n -\frac{1}{D} \int_{S_r} dS(\ve[r]') f_{n-1}(\ve[r'])G_N(\ve[r]\vert\ve[r]')\label{RecRel}
\end{align}
with the condition for $n\ge1$:
\begin{align}
\int_{S_r} dS f_n=0.
\end{align}

For $d=2$, the explicit computations can be done for the first orders, and we find 
\begin{align}
c_0=1/2; \ c_1=\frac{3-\ln4}{2 \pi }, \ c_2=\frac{2}{9}-\frac{7}{3 \pi ^2}.
\end{align}
In the 3D situation, the leading order is simply
\begin{align}
f_0=c_0=1/\pi ; 
\end{align}
and the recurrence relation is 
\begin{align}
f_n(x)=c_n - \frac{2}{\pi}\int_{0}^1 dy f_{n-1}(y)\frac{y}{x+y}K\left(\frac{2\sqrt{x y }}{x+y}\right) .
\end{align}
Unfortunately, it seems very difficult to calculate these integrals, and even at first order the coefficient $c_1$ can be calculated only numerically:
\begin{align}
c_1=\frac{4}{\pi^2}\int_0^1 dx\int_0^1 dy' \frac{xy}{x+y} K\left(\frac{2\sqrt{x y}}{x+y}\right) \simeq 0.27.
\end{align}

\section{Comparison with the constant flux approximation}
\label{CFASection}
\begin{figure}[h!]
\includegraphics[width=8cm,clip]{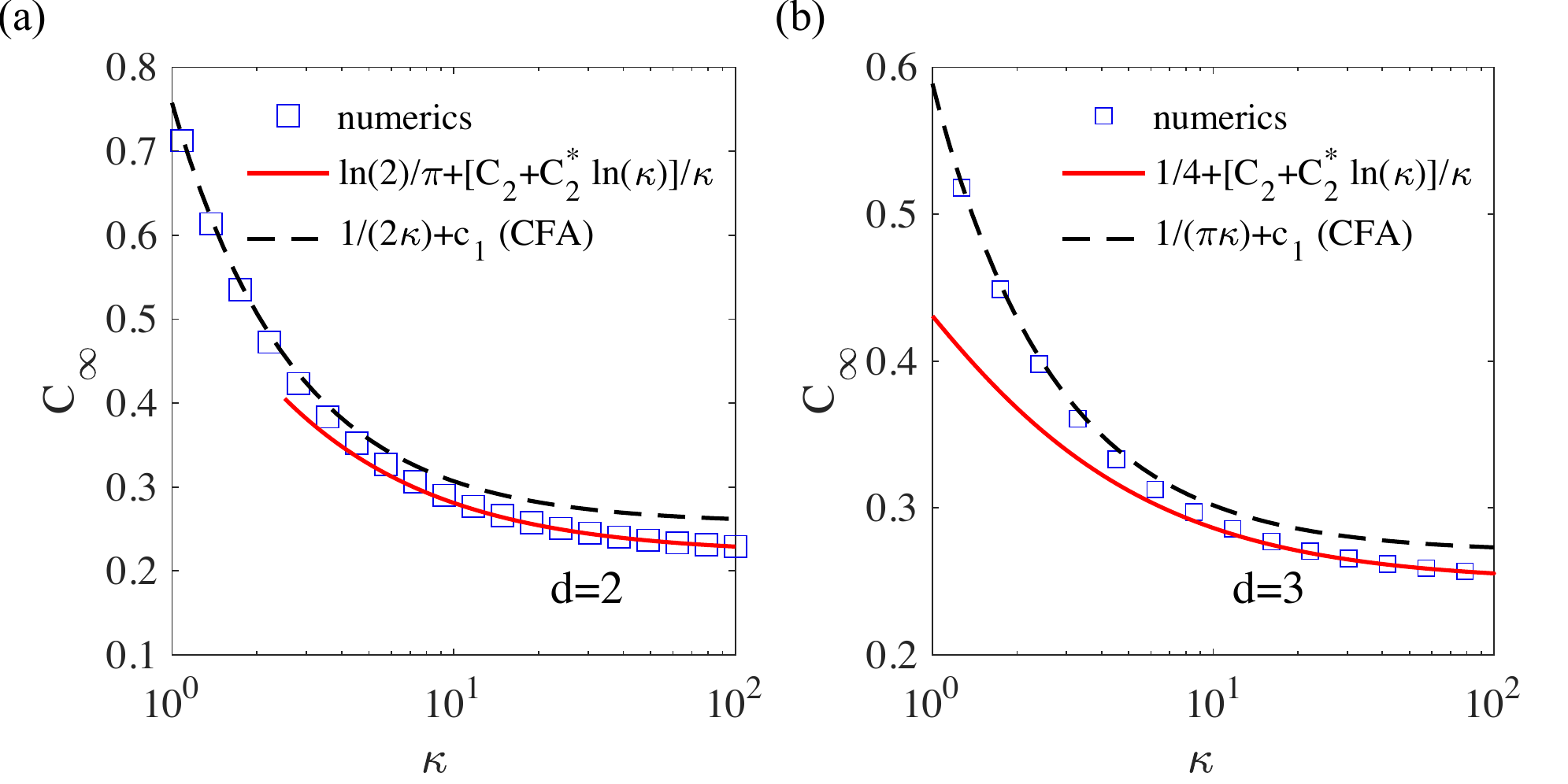}   
\caption{Values of $C_\infty$ in 2D (a) and 3D (b), as found from the exact numerical solution compared to the exact large and small reactivity asymptotics. Note that the constant flux approximation (CFA) is exactly equivalent to the first order expansion in the limit of low reactivity. Here we use the units so that $D=1$, $a=1$. }
\label{Fig_NET_Cinfty}
\end{figure}

The constant flux approximation (CFA) \cite{shoup1981diffusion} has been used in many recent studies  \cite{grebenkov2017diffusive,grebenkov2019full,grebenkov2018towards} on imperfect reactivity in confinement, and here we consider how this approximation compares to the exact results in our formalism. First, we need to adapt this approximation to our situation of large volume limit. 
In the CFA, one replaces the Robin condition (\ref{EqPhiB}) by inhomogeneous \mx{Neumann} conditions:
\begin{align}
& D \partial_z \Phi= 
\begin{cases}
0&(z=0,r>a)\\
- Q &(z=0,r<a)
\end{cases}
\label{CFA}, 
\end{align}
where the flux $Q$ is assumed to be constant on the reactive patch and will be determined self-consistently with a closure relation. A natural choice of closure  relation is to impose that Robin condition is satisfied on average, hence
\begin{align}
Q=\kappa \int_{S_r} dS\   \Phi_s \label{ClosureApprox},
\end{align}
but we also have the normalization condition (\ref{NormCond}), so that 
\begin{align}
Q=1 \label{ValueQ}.
\end{align}
Now, inserting (\ref{CFA}) into  (\ref{89542}) leads directly to a solution for $\Phi$ within CFA:
 \begin{align}
\Phi(\ve[r]_0)=C_\infty+\frac{Q}{D} \int_{S_r} dS(\ve[r]_s)  G_N(\ve[r]_s\vert \ve[r]_0) \label{Phi_CFA}.
\end{align}
 Integrating over $S$ and using (\ref{ValueQ}) and (\ref{ClosureApprox}), we obtain the CFA value of $C_\infty$:
\begin{align}
C_\infty^{\text{cfa}}=\frac{1}{\kappa S_r}\left(1-\frac{\kappa}{D}\int_{S_r}dS(\ve[r])\int_{S_r}dS(\ve[r]_0) G_N(\ve[r]\vert\ve[r]_0)\right)
\end{align} 
Comparing with the results of Sec.~\ref{SmallKappaSect}, we see that in the CFA, $C_\infty$ is exactly the same as the next-to-leading order expansion of $C_\infty$ in the limit of low reactivity, i.e., 
\begin{align}
C_\infty^{\text{cfa}}= \frac{c_0}{\kappa}+c_1 
\end{align}
It may be therefore surprising that CFA works for $C_\infty$ even for rather large values of the reactivity (Fig.~\ref{Fig_NET_Cinfty}), but this comes from the fact that the value of $c_1$ turns out to be extremely close to the exact value of $C_\infty(\kappa=\infty)$ (the difference is of the order of a few percents). This might be the reason why the CFA approach can be implemented to yield accurate results in other contexts. However, the value of the mean first passage time near the extremity of the reactive patch is not well captured by this approximation, since it is obvious in Eq. (\ref{Phi_CFA}) that it does not scale as $1/\sqrt{\kappa}$ for large $\kappa$, contrary to what we have found. 

\section{Conclusion}

In this paper we have considered the imperfect narrow escape problem for diffusive particles in confinement. We have established a general formalism which provides the mean reaction time in  the large volume limit for any value of the reactivity parameter. We have obtained explicit results in $d=2$ and $d=3$ in the respective limits of low and large reactivity parameter. Our most surprising result is the scaling of the mean reaction time when the initial position is at the extremity of the imperfect patch; this mean return time scales as $\kappa^{-1/2}$ and is thus much larger than the naively expected scaling $1/\kappa$. \mx{Interestingly, we have shown that this anomalous scaling is closely related to the divergence of the electric field near corners and edges of conducting objects \cite{jackson1977classical}, which is also responsible for the existence of coffee rings \cite{deegan1997capillary} or the crispiness of the extremities of cooked potatoes \cite{bocquet2007tasting}. } We have explicitly identified \mx{the prefactor of this scaling law } by solving a Wiener-Hopf equation. We have also identified a non-analytic behavior for the capacitances of the imperfect patches as a function of the reactivity.  \mx{We note that we have restricted ourselves to the case of circular patches, but we believe that for the more general patches with a smooth boundary the asymptotic scaling laws should remain unchanged.}
Finally, we have made a link between the results obtained within the Constant Flux Approximation (CFA) and the low reactivity limit. It turns out that the \mx{CFA} gives very accurate predictions of the mean reaction time when the initial position is far from the target, but fails to predict the correct behavior of the mean return times; when the initial position is on the reactive patch. \mx{In the future, one could adapt our formalism to multiple targets, for example to generalize the classical calculation \cite{berg1977physics} of the absorption time by a sphere covered by reactive patches to imperfect patches. Our}  results   provide a general framework to quantify  the mean reaction times for the imperfect narrow escape problem.

  \begin{acknowledgments}
Computer time for this study was provided by the computing facilities MCIA (Mesocentre de Calcul Intensif Aquitain) of the Universit\'e de Bordeaux and of the Universit\'e de Pau et des Pays de l’Adour.   T.~G. acknowledges the support of the grant \textit{ComplexEncounters}, ANR-21-CE30-0020.
\end{acknowledgments}

\appendix
\section{Exact general form of the solution $\Phi(\ve[r])$}
\label{AppendixGeneralFormSolution}

\subsection{Imperfect narrow escape problem in 2D}
Here we describe a way to obtain the general solution of the problem  formed by Eqs.~(\ref{EqPhi})-(\ref{EqPhiB}). It consists in writing the equations in a set of orthogonal coordinates and using the standard method of separation of variables. We first describe this approach in the 2D case, for which we use the elliptic coordinates $\mu,\nu$ defined as 
\begin{align}
x=a\ \mathrm{ch}(\mu)\cos(\nu), \ 
y=a\ \mathrm{sh}(\mu)\sin(\nu).
\end{align}
We calculate the scale factors $h_i=\vert\partial_i\ve[r]\vert$ with $i\in\{\mu,\nu\}$:
\begin{align}
h_\mu=h_\nu=a\sqrt{ \mathrm{ch}^2(\mu)-\cos^2\nu}.
\end{align}
The Laplace equation satisfied by $\Phi$ and the reflecting boundary conditions outside the reactive patch are written in this coordinate system as
\begin{align}
&\partial_\nu^2\Phi+\partial_\mu^2\Phi=0, \ 
&\partial_\nu\Phi\vert_{\nu=0}=\partial_\nu\Phi\vert_{\nu=\pi/2}=0.
\end{align}
The general solution for these equations can be written by using   the method of separation of variables, which leads to 
\begin{align}
\Phi= B \mu  + \sum_{n=0}^\infty \phi_n\  e^{-2n\mu}\cos(2n\nu)\label{Series2D}.
\end{align}
Furthermore, the normalization condition (\ref{NormCond}) can also be written $ D \int dS \partial_n\Phi=1$ for any surface surrounding the target. Far from the target, this means that $\partial_r\Phi=1/( \pi r D)$. Noting that $\mu\simeq\ln(2r/a)$ for large $r$,  we thus find 
\begin{align}
B=1/( \pi D).
\end{align} 
We also note that the quantity $C_\infty$ is given, in  this mode decomposition, by
\begin{align}
C_\infty= \frac{\ln 2}{\pi D} + \Phi_0. \label{Cinf2D_032}
\end{align}
Finally, the Robin condition at the target surface reads
\begin{align}
D \partial_n \Phi+\kappa\Phi = \left(- \frac{D}{h_\mu}\partial_\mu\Phi+\kappa\Phi \right)_{\mu=0}=0   ,
\end{align}
so that
\begin{align}
&D \partial_\mu\Phi\vert_{\mu=0}= \kappa\  a \ \sin\nu\ \Phi\vert_{\mu=0}.
\end{align}
Using this condition and the form of the general solution (\ref{Series2D}), we find that the coefficients $\phi_n$ are solution of the infinite linear system
\begin{align}
  \pi m \phi_m + \frac{\kappa a }{D}\sum_{n=0}^\infty    A_{mn}\phi_n =\delta_{m,0},\label{LinSystem2D}
\end{align}
which is satisfied for all positive integers $m$, with 
\begin{align}
&A_{nm} =\int_0^\pi d\nu \sin\nu \cos(2m \nu)\cos(2n \nu)\nonumber\\
&=\frac{2[1-4(m^2+n^2)]}{16(m^4+n^4)+1-8(m^2+n^2)-32m^2n^2}.
\end{align}
In practice, this linear system (\ref{LinSystem2D}) can be solved numerically by taking into account only a finite number of modes $N$, and checking that the obtained quantities do not depend on $N$ for large $N$. Note also that $C_\infty$ can be directly calculated by using Eq.~(\ref{Cinf2D_032}). 
 
\subsection{3D case}
This approach can be adapted to the 3D case, for which we use orthogonal coordinates defined as
\begin{align}
&x=a \sqrt{(1+\alpha^2)(1-\beta^2)}\cos\varphi,\\
&y=a \sqrt{(1+\alpha^2)(1-\beta^2)}\sin\varphi,\\
&z=a \alpha \beta, 
\end{align}
where $\varphi$ is the azimuthal angle. Note that $\alpha>0$ and $\beta\in[0;1]$ are related to the standard oblate spheroidal coordinates ($\mu,\nu,\varphi)$ by $\alpha=\sinh(\mu)$ and $\beta=\sin(\nu)$. Inversion formulas read
\begin{align}
&\alpha=\sqrt{\frac{\left(\frac{r}{a}\right)^2-1+\sqrt{1+\left(\frac{r}{a}\right)^4+2 \left(\frac{r}{a}\right)^2 \cos(2\theta) }}{2}}, \\
&\beta=\sqrt{\frac{1-\left(\frac{r}{a}\right)^2+\sqrt{1+\left(\frac{r}{a}\right)^4+2 \left(\frac{r}{a}\right)^2 \cos(2\theta) }}{2}}, 
\end{align}
with $(r,\theta,\varphi)$ the usual spherical coordinates. 
It is useful to calculate the scale factors $h_i=\vert \partial{\ve[r]}/\partial i\vert$, with $i=\{\alpha,\beta,\varphi\}$, 
\begin{align}
&h_\alpha=a \left(\frac{\alpha^2+\beta^2}{1+\alpha^2}\right)^{1/2}, \ h_\beta=a \left(\frac{\alpha^2+\beta^2}{1-\beta^2}\right)^{1/2}, \nonumber \\
&h_\varphi=a   \left[ (1+\alpha^2)(1-\beta^2)\right]^{1/2}.
\end{align}
For axisymmetric functions, the Laplacian reads in this orthogonal coordinates
\begin{align}
\nabla^2\Phi=\frac{1}{h_\alpha h_\beta h_\varphi}\left( \frac{\partial}{\partial \alpha} \frac{h_\beta h_\varphi}{h_\alpha} \frac{\partial \Phi}{\partial \alpha}
+\frac{\partial}{\partial \beta} \frac{h_\alpha h_\varphi}{h_\beta} \frac{\partial \Phi}{\partial \beta},  
\right),
\end{align}
so that $\Phi$ satisfies the equation
\begin{align}
\frac{\partial}{\partial \alpha} (1+\alpha^2) \frac{\partial \Phi}{\partial \alpha}+\frac{\partial}{\partial \beta} (1-\beta^2) \frac{\partial \Phi}{\partial \beta}=0. 
\end{align}
We impose \mx{Neumann} conditions for $\beta=0$ and $\beta=1$, at which $\partial_\beta\Phi=0$. With these conditions, 
the general solution  can be found by the method of separation of variables, which leads to
\begin{align}
\Phi(\alpha,\beta)=\Phi_\infty+\sum_{q=0}^\infty a_q \ g_q(\alpha)\ P_{2q}(\beta), \label{DecompositionNET3D}
\end{align}
where $P_{2q}$ are even Legendre polynomials (satisfying both \mx{Neumann} conditions at $\beta=0$ and $\beta=1$), and 
\begin{align}
g_q(\alpha)=\frac{1}{i}Q_{2q}(i  \alpha)-\frac{\pi}{2}P_{2q}(i  \alpha )\label{RobinAlphaBeta}, 
\end{align}
where $i^2=-1$ and $Q_{2q}$ are Legendre functions of the second kind. Let us give here additional details on the function $g_q$. 
To see that $g_q$ is real it is useful to write $Q_{2q}$ as \cite{Abramowitz1964}
\begin{align}
Q_{2q}(x)=\frac{P_{2q}(x)}{2}\ln\frac{1+x}{1-x}-W_{2q-1}(x),
\end{align}
where $W$ is the polynomial
\begin{align}
W_{2q-1}(x)=\sum_{m=1}^q\frac{4q-(1+4(m-1)) }{(2m-1)(2q-m+1)}P_{2q-(2m-1)}(x).
\end{align}
For purely imaginary arguments $x=i \alpha$, we have 
\begin{align}
Q_{2q}(i\alpha)= i P_{2q}(i \alpha)\arctan(x)-W_{2q-1}(i\alpha),
\end{align}
and we thus see that 
\begin{align}
g_q(\alpha)=P_{2q}(i \alpha)\arctan(\alpha)-\frac{W_{2q-1}(i\alpha)}{i}-\frac{\pi}{2}P_{2q}(i \alpha ). 
\end{align}
Using the parity of $P$ and $W$, it becomes clear that $g_q$ is real. Furthermore it can be checked that it decreases to zero at infinity (and $g_0\sim 1/\alpha$ for large $\alpha$). 

Now, the equation satisfied by the coefficients $a_q$ is identified by using the Robin condition. In these coordinates, the partially absorbing disk  corresponds to $\alpha=0$, and the Robin conditions can be deduced from $\partial_n\Phi=-\left(h_\alpha^{-1}\partial_\alpha\Phi\right)_{\alpha=0} $ so that the boundary conditions read
\begin{align}
(D \partial_\alpha\Phi- a \beta \kappa \Phi)_{\alpha=0} =0 .
\end{align}
Inserting the general solution  (\ref{DecompositionNET3D}) into the above boundary condition, multiplying by $P_{2k}(\beta)$ and integrating, we obtain the linear system:
\begin{align}
&\sum_{q=0}^{\infty} a_q g_q'(0)\int_0^1 d\beta P_{2q}(\beta)P_{2k}(\beta)=\kappa\Phi_\infty\int_0^1d\beta \beta P_{2k}(\beta) \nonumber\\
&+\kappa\sum_{q=0}^\infty a_q g_q(0)\int_0^1 d\beta \beta P_{2q}(\beta)P_{2k}(\beta)\label{094321},
\end{align}
for all positive integers $k$. Finally, we calculate the surface element at $\alpha=0$, $dS_\alpha = h_\varphi h_\beta d\varphi d\beta$ so that the normalization condition reads
\begin{align}
2\pi \kappa a^2 \int_0^1 \Phi(0,\beta)\beta d\beta=1,  \label{NormOrthog} 
\end{align}
which leads to the equation
\begin{align}
\kappa a^2\pi \Phi_\infty + 2\pi \kappa a^2  \sum_{q=0}^\infty a_q g_q(0) \int_0^1 d\beta\beta P_{2q}(\beta)=1.
\end{align}
A numerical solution for $\Phi_\infty$ can thus be found by solving the linear system (\ref{094321}) for the coefficients $a_q$ (completed by the above normalisation condition). Note also that $C_\infty=\Phi_\infty$.

\section{Identification of the integral equation (\ref{IntEq3D}) in 3D}
\label{AppendixEqInt3D}
Here we briefly show how to obtain the integral equation (\ref{IntEq3D}). Using Eq.~(\ref{GreenFunction}) for $d=3$, we see that Eq.~(\ref{IntEqForPhiAnyD}) writes 
\begin{align}
\Phi(r_0)=C_\infty -  \frac{\kappa}{D} \int_0^a dr K(r,r_0)\Phi(r) ,\label{7432}
\end{align}
with 
\begin{align}
K(r,r_0)&=\frac{1}{2\pi}\int_0^{2\pi}\frac{r \ d\theta}{\sqrt{r^2+r_0^2-2 r r_0 \cos\theta}} .
\end{align}
The quantity $K(r,r_0)$ can be simplified  as follows:
\begin{align}
K(r,r_0) &=\frac{1}{ \pi}\int_0^{\pi}\frac{r \ d\theta}{\sqrt{r^2+r_0^2-2 r r_0 \cos\theta}}\nonumber\\
&= \frac{2r}{ \pi}\int_0^{\pi/2}\frac{du}{\sqrt{r^2+r_0^2-2 r r_0 \cos(2u)}}\nonumber\\
&= \frac{2r}{ \pi }\int_0^{\pi/2}\frac{du}{\sqrt{r^2+r_0^2-2 r r_0 [2\cos^2u-1]}}\nonumber\\
&=\frac{2r}{ \pi (r+r_0)} \int_0^{\pi/2}\frac{du}{\sqrt{1-\frac{4 r r_0}{(r+r_0)^2}\cos^2u}}\nonumber\\
&= \frac{2r}{ \pi (r+r_0)} K\left(\frac{2 \sqrt{r r_0}}{r+r_0}\right),
\end{align}
where $u=\theta/2$ and in the last line we have recognized the definition of the elliptic function $K$. Inserting the above result into Eq.~(\ref{7432}) finally leads to Eq.~(\ref{IntEq3D}).

\section{Calculation details for the identification of $C_2$ and $C_2^*$}
\subsection{2D situation}
\label{CalculationC2_2D}

Here we describe the details of calculations leading to the identification of the constants $C_2$ and $C_2^*$ in the 2D situation. Let us first evaluate the term $B$ in Eq.~(\ref{DefB}): 
\begin{align}
&B\simeq   - \frac{ 1}{\sqrt{\kappa} }\int_{0}^{\ell\kappa} dX  \left(\psi(X)-\frac{1}{\pi\sqrt{2X}}\right) \ln ( 1 - x_0 ) \nonumber \\
&-\int_0^{\ell\kappa} \frac{ dX}{\sqrt{\kappa} }  \left(\psi(X)-\frac{1}{\pi\sqrt{2X}}\right)\ln\frac{ 1 - x_0 -X/\kappa  }{1-x_0}  .
\end{align}
Here we have only assumed that one can use the leading order of the scaling form for the solution $\Phi$ for arguments lower than $\ell$. Now, to evaluate the first line we use the normalisation condition (\ref{Cond}), and to evaluate the terms on the second line, we change variable $X=u \kappa\ell$:
\begin{align}
&B\simeq  \frac{ 1}{\sqrt{\kappa} }\int_{\ell\kappa}^\infty dX  \left(\psi(X)-\frac{1}{\pi\sqrt{2X}}\right) \ln ( 1 - x_0 ) \nonumber \\
&- \sqrt{\kappa}\ell \int_0^{1} du  \left(\psi(\kappa u\ell)-\frac{1}{\pi\sqrt{2\kappa u\ell }}\right) \ln \frac{ 1 - x_0 -u \ell }{1-x_0}. 
\end{align}
Using Eq.~(\ref{PsiLargeX}),   we can write $B$ under the form
\begin{align}
B&\simeq \frac{f_2^*(x_0) \ln\kappa+f_2(x_0)}{\kappa} ,
\end{align}
with
\begin{align}
f_2^*(x_0)&= \frac{ \ln   (1 - x_0   ) }{   \sqrt{\ell}  \sqrt{2} \pi ^2  }   
-  \int_0^{1}  \frac{du \ \ell}{\pi^2(2u\ell)^{\frac{3}{2}}} \ln \frac{ 1 - x_0-u\ell }{1-x_0},\label{Def_f2} \\
f_2(x_0)&=
 \frac{\ln ( 4\ell)+\gamma_e +1}{\sqrt{2\ell} \pi ^2  }  \ln   (1 - x_0   )\nonumber\\
&- \ell\int_0^{1} du \frac{ \ln (4 u\ell  )+\gamma_e -1}{\pi^2(2u\ell)^{3/2}} \ln \frac{ 1 - x_0-u\ell }{1-x_0} .
 \end{align}
Let us precise a few properties of $f_2^*$ (similar properties hold for $f_2$). In the limit $\ell\to0$ at fixed $x_0$, we see that 
\begin{align}
f_2^*(x_0)&\underset{\ell\to0}{\sim} \frac{ \ln   (1 - x_0   ) }{   \sqrt{\ell}  \sqrt{2} \pi ^2  }   \label{95432143}
\end{align} 
Near the extremity of the patch, we set $x_0=1-v\ell$ to determine the behavior of $f_2^*$. In the limit $\ell\to0$ at fixed $v=(1-x_0)/\ell$, we obtain
\begin{align}
f_2^*(x_0=1-v\ell)&\underset{\ell\to0}{\sim} \frac{ \ln   (v\ell   ) }{   \sqrt{\ell}  \sqrt{2} \pi ^2  }   
-  \int_0^{1}  \frac{du \ \ell}{\pi^2(2u\ell)^{\frac{3}{2}}} \ln \frac{ v-u  }{v}  \label{05421954}
\end{align} 

Collecting the terms $O(\ln\kappa/\kappa)$ in the integral equation (\ref{TruncatedIntegralEq}) leads to
\begin{align}
&\int_{-1+\ell}^{1-\ell} dx   \  \Phi_2^*(x)   \ln\vert x-x_0\vert   = F_2^*(x_0),  \label{EqIntergralTruncated}  \\
&F_2^*(x_0)= -\pi C_2^* + f_2^*(x_0)+f_2^*(-x_0).\label{DefBigF2Star}
\end{align}
We consider this equation as an integral equation over the interval $[-1+\ell; 1-\ell]$, for which the solution is analytically known \cite{HandbookIntegralEquations}: 
\begin{align}
\Phi_2^*(x)=\frac{1}{\pi^2 \sqrt{b^2-x^2}}\Bigg[\int_{-b}^{b} dt \frac{\sqrt{b^2-t^2} \partial_t F_2^*(t)}{t-x} \nonumber\\+\frac{1}{\ln[b/2]}\int_{-b}^{b} dt \frac{F_2^*(t)}{\sqrt{b^2-t^2}}\Bigg],
\end{align}
where we have set $b=1-\ell$. As a consequence, the integral of $\Phi_2^*$ over the truncated interval $]-b;b[$ reads
\begin{align}
I_2^*(\ell)&=\int_{-b}^{b} dx\ \Phi_2^*(x) = \frac{1}{\pi \ln[b/2]}\int_{-b}^{b}  \frac{ dx\ F_2^*(x)}{\sqrt{b^2-x^2}}.
\end{align} 
When $\ell\to0$, the behavior of $I_2^*(\ell)$ is obtained by inserting the small $\ell$ limit of $f_2^*(x_0)$ at fixed $x_0$ given by Eq.~(\ref{95432143}) into Eq.~(\ref{DefBigF2Star}) and inserting the result into the above equation, leading to
\begin{align}
I_2^*(\ell)\underset{\ell\to0}{\simeq}   \frac{(-1)}{\pi \ln2}\int_{-1}^{1}  \frac{dt\ \ln \left(1-t^2\right)}{\sqrt{2} \pi ^2\sqrt{\ell}\sqrt{1-t^2}} =\frac{\sqrt{2}}{\pi^2 \sqrt{\ell}},
 \end{align}
 this result is consistent with the fact that the behavior of $\Phi_2^*$ near $x=\pm1$ is given by (\ref{MatchCond2Star}). Now, the fact that the normalisation condition (\ref{NormCond2D}) holds at all powers of $\kappa$ implies that
 \begin{align}
	\lim_{\ell\to0} \left[I_2^*(\ell)-\frac{\sqrt{2}}{\pi^2 \sqrt{\ell}}\right]=0.
 \end{align}
We thus evaluate 
\begin{align}
&\Delta I_2^*(\ell) = I_2^*(\ell)-\frac{\sqrt{2}}{\pi^2 \sqrt{\ell}}  \nonumber\\
&\simeq  \frac{(-1)}{\pi \ln2} \int_{-1}^{1} dt\left[\frac{F_2^*(t)\theta(b -\vert t\vert)}{\sqrt{b^2-t^2}} - \frac{\ \ln \left(1-t^2\right)}{ \pi ^2 \sqrt{2\ell(1-t^2)}} \right].
\end{align}
The contributions of $f_2^*(t)$ in this integral are negligible except for $x_0$ at the vicinity of $1$. Thus, we set $t=1-v\ell$ to evaluate the above integral, and the integral can be evaluated by integrating  $v$ over $[0,\infty]$ (except for the term $C_2^*$ which is evaluated without this change of variable). Using (\ref{05421954}) yields
\begin{align}
&\Delta I_2^*(\ell) \simeq  \frac{(-1)}{\pi \ln2} \Bigg[\int_{-1}^{1} dt\frac{-\pi C_2^*}{\sqrt{1-t^2}} \nonumber\\
&+\frac{2}{ \pi ^2 \sqrt{2}}  \int_0^\infty dv \ln(2v\ell) \left(\frac{\theta(v-1)}{\sqrt{2(v-1)}}-\frac{1}{\sqrt{2v}}\right)\nonumber\\
&-  \int_1^\infty dv   \int_0^{1} d{u} \frac{ 2}{\pi^2(2 {u})^{3/2}\sqrt{2(v-1)}} \ln \frac{ v- {u} }{v} . \Bigg]
\end{align}
All the integrals appearing in the above equations can be evaluated. Imposing $\Delta I_2^*(\ell) =0$ then leads to
\begin{align}
C_2^*=1/\pi^2.
\end{align}

To identify $C_2$ we proceed in the same way.  The integral equation is
\begin{align}
&\int_{-1+\ell}^{1-\ell} dx   \  \Phi_2(x)   \ln\vert x-x_0\vert   = F_2(x),\\
&F_2(x)= -\pi C_2  + \pi \phi_1  +f_2(x_0) +f_2(-x_0),
\end{align}
so that 
\begin{align}
I_2(\ell)=\int_{-b}^{b}dx\ \Phi_2(x) = \frac{1}{\pi \ln(b/2)}\int_{-b}^{b} \frac{dt\ F_2(t)}{\sqrt{b^2-t^2}}.
\end{align}
As before we need to evaluate the behavior of $F_2(x,\ell)$ for small $\ell$
\begin{align}
&F_2(x)\underset{\ell\to0}{\sim} 
 \frac{\ln ( 4\ell)+\gamma +1}{\sqrt{2\ell} \pi ^2  }  \ln   (1 - x_0^2   )  = \frac{F_2^0(x)}{\sqrt{\ell} }.
  \end{align}
At leading order for small $\ell$ we obtain
\begin{align}
I_2(\ell)\underset{\ell\to0}{\sim} \frac{1}{\pi \ln(1/2)}\int_{-1}^{1}dt \frac{ F_2^0(t)}{\sqrt{\ell}\sqrt{1-t^2}}=\frac{I_2^0}{\sqrt{\ell}}.
\end{align}
The next-to-leading order is
\begin{align}
I_2-\frac{I_2^0}{\sqrt{\ell}} \simeq \frac{(-1)}{\pi \ln2} \int_{-1}^{1} dt\left[\frac{F_2(t)\theta(b -\vert t\vert)}{\sqrt{b^2-t^2}} - \frac{F_2^0(t)}{  \sqrt{\ell(1-t^2)}}\right] 
\end{align}
Again, we evaluate it by setting $t=1-v \ell$ and taking the small $\ell$ limit of the obtained integrand at fixed $v$, leading to
\begin{align}
&I_2(\ell)-I_2^0(\ell) \simeq \frac{(-1)}{\pi \ln2} \Bigg\{ \int_{-b}^{b}dx \frac{-\pi C_2+\Phi_1(x)}{\sqrt{b^2-x^2}} \nonumber\\
&+2\int_{0}^{\infty} dv  \frac{\ln ( 4\ell)+\gamma +1}{ \sqrt{2 } \pi ^2  }  \ln  (2 v \ell )  \left[ \frac{\theta(v-1)}{\sqrt{2(v-1)}}-\frac{1}{\sqrt{2v}}\right] 
\nonumber\\
&-2\int_1^\infty dv \int_0^\infty du  \frac{ \ln (4  {u}\ell  )+\gamma_e -1}{ \pi^2(2 {u})^{3/2} \sqrt{2(v-1)}}  \ln \frac{ v- {u} }{v}\Bigg\} \label{0543811}
\end{align}
To evaluate the term involving $\Phi_1$ we introduce a variable $\varepsilon$ so that $\ell \ll \varepsilon\ll 1$ and we calculate
\begin{align}
 \int_{-b}^{b}dx& \frac{\Phi_1(x)}{\sqrt{b^2-x^2}}=\int_{-b}^{b} \frac{dx}{\pi\sqrt{1-x^2}\sqrt{b^2-x^2}}\nonumber\\
  &\simeq \int_{-1+\varepsilon}^{1-\varepsilon}  \frac{dx}{\pi(1-x^2)}+\int_1^{\varepsilon/\ell}    \frac{2\ell \ dv}{2\pi\sqrt{v(v-1}}\nonumber\\
  &\simeq \frac{\ln(8/\ell)}{\pi},
\end{align}
where the last equality follows from the evaluation of the integrals with  $\ell \ll \varepsilon\ll 1$. 
Finally, evaluating all terms in Eq.~(\ref{0543811}) and requiring that the $I_2(\ell)-I_2^0(\ell)\to0$ for small $\ell$, we obtain
\begin{align}
 C_2=\frac{1+\gamma_e+\ln 8}{\pi^2},
\end{align} 
as announced in the main text. 

\subsection{Second-order calculation in the limit of large reactivity in 3D}
We evaluate the term $B$ in Eq.~(\ref{DefB3D}) by writing
\begin{align}
&B(r_0,\ell)     \simeq  \sqrt{\kappa}    \frac{K\left(\frac{2\sqrt{  r_0}}{1+r_0} \right)}{(1+r_0)} \int_{\ell\kappa}^\infty dX    \left[\psi(X)-\frac{1}{ 2\pi \sqrt{2X } }\right] \nonumber\\
& -\ell \int_{0}^1 \frac{du}{\sqrt{\kappa}(1+r_0)}
 \left[\psi(u\ell\kappa)-\frac{1}{2\pi \sqrt{2u\ell\kappa}}\right] \nonumber\\
&\times\left[K\left(\frac{2\sqrt{(1-u\ell) r_0}}{1-u\ell+r_0} \right)- K\left(\frac{2\sqrt{r_0}}{1+r_0} \right) \right].
 \end{align}
Hence
\begin{align}
&B(r_0,\ell)=  \frac{K\left(\frac{2\sqrt{  r_0}}{1+r_0} \right) }{(1+r_0)} \frac{1+\gamma_e+\ln(4\ell\kappa)}{2\sqrt{2\ell}\pi^2}\nonumber\\
&  -\ell  \int_{0}^1 du\frac{[-1+\gamma_e+\ln(4u\ell\kappa)]}{2 \pi^2 (2u\ell)^{3/2}(1+r_0)}  \nonumber\\
& \times  \left[ K\left(\frac{2\sqrt{(1-u\ell) r_0}}{1-u\ell+r_0} \right)- K\left(\frac{2\sqrt{r_0}}{1+r_0} \right) \right].
\end{align}
Note that here, for conciseness we will treat $\ln \kappa$ as being of order $1$ in powers of $\kappa$, the result will be exactly the same as in the case where one separates the $\ln\kappa$ terms and the $O(1)$ terms. 

 In the small $\ell$ limit at fixed $r_0$ we obtain
\begin{align}
B(r_0,\ell)\underset{\ell\to0}{\sim}  \frac{K\left(\frac{2\sqrt{  r_0}}{1+r_0} \right)}{1+r_0}  \frac{1+\gamma_e+\ln(4\ell\kappa)}{2\sqrt{2\ell}\pi^2}
=\frac{B^{0}(r_0)}{\sqrt{\ell}}, 
\end{align}
whereas if we set $r_0=1-\ell v$,  in the limit $\ell\to0$ at fixed $v$ we obtain
\begin{align}
B(1-v\ell,\ell)\underset{\ell\to0}{\sim} \frac{1}{8}\frac{1+\gamma_e+\ln(4\ell \kappa)}{\sqrt{2\ell}\pi^2}\ln\frac{8^2}{(v\ell)^2} \nonumber \\
-  \int_{0}^1 d {u} \ 
\frac{[-1+\gamma_e+\ln(4\ell \kappa  {u})]}{4 \pi^2 (2 {u})^{3/2}}  \ln\frac{v}{ v- {u} }  ,\label{B29}
\end{align}
where we have used $K(1-y)\simeq \frac{1}{2}\ln(8/y)$ for small $y$. 
Let us write the integral equation (\ref{5843}) under the form 
\begin{align}
\int_0^{1-\ell}  \ \frac{dr\ r \ \tilde{\Phi}_2(r)}{r+r_0}K\left(\frac{2\sqrt{r r_0}}{r+r_0} \right)= 
 \frac{\pi [\tilde{C}_2-\Phi_1]}{2} +B\label{IntEqPhi2Tilde}
\end{align}
with $\tilde{\Phi}_2=\Phi_2(r)+\Phi_2^*(r)\ln\kappa$, $\tilde{C}_2=C_2^*\ln\kappa+C_2$. 
Let us define
\begin{align}
I_2(r,\ell)=\int_0^{1-\ell} dr \ r \ \tilde{\Phi}_2(r).
\end{align}
Using the analytically known solution \cite{HandbookIntegralEquations} of the integral equation (\ref{IntEqPhi2Tilde}), we obtain  
\begin{align}
I_2(\ell)=\frac{4}{\pi^2} \int_0^{1-\ell} ds\frac{s \left(\frac{\pi }{2}(C_2 -\Phi_1)+B(s,\ell)\right) }{\sqrt{(1-\ell)^2-s^2}} .
\end{align}
When $\ell\to0$ we obtain at leading order
\begin{align}
I_2(r,\ell)\underset{\ell\to0}{\sim}\frac{4}{\pi^2\sqrt{\ell}} \int_0^{1} ds \frac{s }{\sqrt{1-s^2}}  B^{0}(s,\ell)
\end{align}
and this integral diverges for $\ell\to0$, as it should due to the known behavior for $\tilde{\Phi}_2(r)$ when $r$ approaches $1$.  
At next-to-leading order, we evaluate the terms involving $B$ by setting $s=1-v\ell$ and take the small $\ell$ limit at fixed $v$, so that we can use Eq.~(\ref{B29}):
 \begin{align}
&I_2(\ell)-\frac{4}{\pi^2\sqrt{\ell}} \int_0^{1}  \frac{ds \ s B^{0}}{\sqrt{1-s^2}}  
 \simeq \frac{4}{\pi^2}\Bigg\{  \int_0^{1} \frac{ds \ s }{\sqrt{1-s^2}} \frac{\pi }{2}C_2 \nonumber\\
& - \int_1^\infty dv \int_{0}^1 d {u} \ 
\frac{[-1+\gamma_e+\ln(4\ell\kappa {u})]}{4 \pi^2 (2u)^{3/2}\sqrt{2(v-1)}}  \ln\frac{v}{v- {u}}   \nonumber\\
&+\int_0^\infty dv \frac{1+\gamma_e+\ln(4\ell\kappa)}{4\sqrt{2 }\pi^2}\ln\frac{8}{v\ell} \left(\frac{\theta(v-1)}{\sqrt{2(v-1)}}-\frac{1}{\sqrt{2v}}\right)\nonumber\\
&+ \int_0^{1-\ell} ds\frac{s \pi \Phi_1(s) }{2\sqrt{(1-\ell)^2-s^2}} \label{04592144}
\Bigg\}.
\end{align}
To evaluate the term containing $\Phi_1$, defined in Eq.~\eqref{954230} we can use again a trick where we use a variable $\varepsilon$ with $\ell \ll \varepsilon\ll 1$:
\begin{align}
\int_{0}^{1-\ell} \frac{dr \ r \ \Phi_1(r)}{\sqrt{(1-\ell)^2-r^2}}=\int_{0}^{1-\varepsilon} dr \frac{r \Phi_1(r)}{\sqrt{1-r^2}}\nonumber\\
+ \int_1^{\varepsilon/\ell}    \frac{dv\ \ell}{4\pi\sqrt{v(v-1}}=\frac{\ln(2/\ell)}{4\pi}.
\end{align}
Finally, all the integrals in (\ref{04592144}) can be evaluated, requiring that it vanishes for small $\ell $ leads to 
\begin{align}
\tilde{C}_2= \frac{\gamma_e +1+\ln (2\kappa)}{4 \pi },
\end{align}
which is exactly Eq.~(\ref{ValueC2_3D}). 
 



\end{document}